\documentclass[aps,pra,reprint,floatfix,nofootinbib,superscriptaddress,longbibliography]{revtex4-1}

\usepackage{amsmath,amsthm,amsfonts,amssymb,bm}
\usepackage{graphicx,epstopdf,nomencl,upgreek,url}
\usepackage{times}
\usepackage[colorlinks={true},pdftex]{hyperref}  
\usepackage{subfigure}
\usepackage{color,xcolor,soul}
\usepackage{braket}
\usepackage{tabularx}

\definecolor{LY}{rgb}{1.0,1.0,0.7}
\sethlcolor{LY}

\hypersetup{
  colorlinks,
  citecolor=blue,
  linkcolor=blue,
  urlcolor=blue}
  
 \AtBeginDocument{%
    \newwrite\bibnotes
    \def\bibnotesext{Notes.bib}
    \immediate\openout\bibnotes=\jobname\bibnotesext
    \immediate\write\bibnotes{@CONTROL{REVTEX41Control}}
    \immediate\write\bibnotes{@CONTROL{%
    apsrev41Control,author="08",editor="1",pages="1",title="0",year="1"}}
     \if@filesw
     \immediate\write\@auxout{\string\citation{apsrev41Control}}%
    \fi
}%
  
\begin{document}

\title{Characterization of Errors in Interferometry with Entangled Atoms}
\author{Constantin Brif}
\email[Corresponding author.~]{cnbrif@sandia.gov}
\affiliation{Sandia National Laboratories, Albuquerque, NM 87185 and Livermore, CA 94550, USA}
\author{Brandon P. Ruzic}
\email{bruzic@sandia.gov}
\affiliation{Sandia National Laboratories, Albuquerque, NM 87185 and Livermore, CA 94550, USA}
\author{Grant W. Biedermann}
\email{biedermann@ou.edu}
\affiliation{Sandia National Laboratories, Albuquerque, NM 87185 and Livermore, CA 94550, USA}
\affiliation{Department of Physics and Astronomy, University of Oklahoma, Norman, OK 73019, USA}

\begin{abstract}
Recent progress in generating entangled spin states of neutral atoms provides opportunities to advance quantum sensing technology. In particular, entanglement can enhance the performance of accelerometers and gravimeters based on light-pulse atom interferometry. We study the effects of error sources that may limit the sensitivity of such devices, including errors in the preparation of the initial entangled state, imperfections in the laser pulses, momentum spread of the initial atomic wave packet, measurement errors, spontaneous emission, and atom loss. We determine that, for each of these errors, the expectation value of the parity operator $\Pi$ has the general form, $\overline{\braket{\Pi}} = \Pi_0 \cos( N \phi )$, where $\phi$ is the interferometer phase and $N$ is the number of atoms prepared in the maximally entangled Greenberger--Horne--Zeilinger state. Correspondingly, the minimum phase uncertainty has the general form, $\Delta\phi = (\Pi_0 N)^{-1}$. Each error manifests itself through a reduction of the amplitude of the parity oscillations, $\Pi_0$, below the ideal value of $\Pi_0 = 1$. For each of the errors, we derive an analytic result that expresses the dependence of $\Pi_0$ on error parameter(s) and $N$, and also obtain a simplified approximate expression valid when the error is small. Based on the performed analysis, entanglement-enhanced atom interferometry appears to be feasible with existing experimental capabilities.
\end{abstract}


\maketitle

\section{Introduction}
\label{sec:intro} 

Neutral atoms are used in some of the most precise, state-of-the-art quantum sensors, including atomic clocks~\cite{Ludlow.RMP.87.637.2015, Heavner.Metrologia.51.174.2014, Bloom.Nature.506.71.2014, Bothwell.Metrologia.56.065004.2019}, optical atomic magnetometers~\cite{Budker.NatPhys.3.227.2007, Budker.book.2013, Wilson.PRAppl.11.044034.2019}, and atom interferometers (AIs)~\cite{Cronin.RMP.81.1051.2009,Wang.CPB.24.053702.2015, Kovachy.Nature.528.530.2015,Bongs.NatRevPhys.1.731.2019}. Looking forward, an ambitious goal is to harness the power of quantum entanglement to decrease the phase uncertainty in these atomic sensors~\cite{Pezze.RMP.90.035005.2018}, first, beyond the standard quantum limit (SQL), $N^{-1/2}$, which arises in measurements with independent atoms, and, ultimately, as close as possible to the Heisenberg limit (HL), $N^{-1}$, where $N$ is the number of atoms used in the measurement.

There are two main approaches to generating metrologically useful entanglement in atomic systems. One approach utilizes spin squeezing \cite{Wineland.PRA.46.R6797.1992, Wineland.PRA.50.67.1994, Kitagawa.PRA.47.5138.1993} generated in an ensemble of ultracold neutral atoms~\cite{Bohnet.NatPhoton.8.731.2014, Cox.PRL.116.093602.2016, Hosten.Nature.529.505.2016, Engelsen.PRL.118.140401.2017}. It has been shown that spin squeezing makes it possible to surpass the SQL in atomic clocks \cite{Hosten.Nature.529.505.2016, Malia.arxiv.1912.10218} and atomic magnetometers \cite{Sewell.PRL.109.253605.2012, Muessel.PRL.113.103004.2014}. In these experiments, spin-squeezed states were generated in large ensembles of atoms ($10^5$--$10^6$), but only a small fraction of the atoms ($\sim 0.1\%$) have been actually entangled~\cite{Hosten.Nature.529.505.2016}. Also, strong spin squeezing is generated via an optical-cavity-based measurement, and releasing the atoms from the cavity into free space (which is typically required in order to use them in an AI) results in a fast degradation of squeezing~\cite{Wu.arxiv.1912.08334}.

An alternative approach is to produce maximally entangled atomic spin states such as the Greenberger--Horne--Zeilinger (GHZ) state \cite{GHZ.arxiv.0712.0921}, which are known to achieve the HL under ideal conditions \cite{Bollinger.PRA.54.R4649.1996}. The most promising method for generating high-fidelity entangled states of atomic spins is by using Rydberg-mediated interactions in arrays of ultracold, optically trapped neutral atoms~\cite{Weiss.PT.70.44.2017}. In particular, this method was used in a number of experiments~\cite{Jau.NatPhys.12.71.2016, Martin.SAND-report.2018, Levine.PRL.121.123603.2018, Graham.PRL.123.230501.2019, Levine.PRL.123.170503.2019} to generate maximally entangled two-spin states (i.e., Bell states \cite{Nielsen.Chuang.2010}), with fidelity as high as 0.97~\cite{Levine.PRL.121.123603.2018}. This method, combined with optimal control~\cite{Brif.ACP.148.1.2012}, also produced GHZ spin states in arrays of 4 to 20 atoms with fidelities of 0.85 to 0.54, respectively \cite{Omran.Science.365.570.2019}. Additional proposed advances such as the use of rapid adiabatic Rydberg dressing~\cite{Mitra.PRA.101.030301.2020} and \emph{in situ} adaptive optimal control~\cite{Kaubruegger.PRL.123.260505.2019}, along with various technical improvements \cite{Omran.Science.365.570.2019}, are likely to further increase the number of entangled atoms and enhance fidelities in these systems. Although maintaining a high fidelity as the number of entangled atoms increases is an outstanding challenge, the possibility of producing highly entangled spin states of ultracold neutral atoms is a promising avenue for quantum sensing.

One of the most prominent sensing techniques for neutral atoms is light-pulse atom interferometry \cite{Borde.PLA.140.10.1989, Kasevich.PRL.67.181.1991, Kasevich.APB.54.321.1992, Young.chapter.1997}. Inertially sensitive AIs have many important practical applications~\cite{Cronin.RMP.81.1051.2009, Wang.CPB.24.053702.2015, Bongs.NatRevPhys.1.731.2019}, including accelerometers and gyroscopes~\cite{Lenef.PRL.78.760.1997, Gustavson.PRL.78.2046.1997, McGuinness.APL.100.011106.2012, Parazzoli.PRL.109.230401.2012, Dickerson.PRL.111.083001.2013}, and gravimeters~\cite{Snadden.PRL.81.971.1998, Fixler.Science.315.74.2007, Dimopoulos.PRD.78.042003.2008, Sorrentino.APL.101.114106.2012}. Promising advances also demonstrate gravity measurements using spatially separated atomic wave packets suspended in a lattice~\cite{Clad_2005, Charriere.PRA.85.013639.2012, Zhang.PRA.94.043608.2016} and most recently with ultralong interrogation times~\cite{Xu.Science.366.745.2019}.

Feeding atoms prepared in a highly entangled spin state into an AI~\cite{Dowling.PRA.57.4736.1998, Borzeszkowski.PLA.286.102.2001, Yurtsever.EPJD.22.365.2003} opens up an exciting path towards inertially sensitive measurements with a phase uncertainty beyond the SQL or even close to the HL. The unparalleled control achieved in ultracold neutral atom experiments~\cite{Weiss.PT.70.44.2017, Omran.Science.365.570.2019} makes them an ideal platform to both reach the fundamental limits of AI-based sensors and understand key errors that affect their performance.

In this paper, we provide a detailed analysis of a light-pulse AI utilizing ultracold atoms that have been prepared in the GHZ spin state. We devise a protocol for the operation of the entanglement-enhanced AI and discuss the relevant physical parameters that affect its performance. In the ideal case of an AI without any errors, the maximally-entangled GHZ state achieves a phase uncertainty at the HL, similarly to the analogous result in spectroscopy \cite{Bollinger.PRA.54.R4649.1996}. Of course, in any physical implementation of the AI, there will be many sources of noise that reduce its sensitivity from the ideal case. Our main goal is to characterize the non-negligible sources of noise and develop mitigation strategies for them if necessary. First, we consider the effect of imperfections in the initial many-atom GHZ state. Next, we investigate how various types of noise affect the operation of the AI after the GHZ state has been prepared with a given fidelity. We characterize the relevant noise sources in terms of error parameters and analyze how the phase uncertainty scales as a function of the number of entangled atoms $N$ in the initial GHZ state. We use the error parameters and the scaling laws we derive to learn about the fundamental limitations of this type of sensor.

The non-negligible errors we consider come from noise in the initial state preparation, laser intensity fluctuations, laser phase noise, the initial momentum spread of the atoms, measurement errors, spontaneous emission during Raman pulses, and atom loss. The initial momentum spread leads to a detuning error in the light pulses of the AI, and we find that this error is dominant for AI operation. This error grows linearly with the atom's vibrational energy in the trap and decreases with the square of the effective Rabi frequency for the two-photon Raman transition. Consequently, the effect of the initial momentum spread can be reduced by cooling the atom to the ground state of the trap~\cite{Kaufman.PRX.2.041014.2012}, by decreasing the trap frequency through adiabatic lowering of the trap depth~\cite{Tuchendler.PRA.78.033425.2008}, and by increasing the Rabi frequency through the use of a high-intensity laser with tight focusing. We forecast that a realistic implementation of an entanglement-enhanced AI can achieve inertial sensing near the HL for $\sim 100$ atoms.

We also develop a detection scheme that allows for a measurement of the interferometric phase $\phi$, with a number of measurements that scales linearly with $N$, in contrast to the exponential scaling of the Hilbert space dimension for a system of $N$ entangled atoms. This is possible because a measurement of the parity of the final $N$-atom spin state, which provides a sufficient amount of information to determine $\phi$, can be performed via a state-selective detection~\cite{Martinez.PRL.119.180503.2017, Kwon.PRL.119.180504.2017, Wu.NatPhys.15.538.2019} with only $N+1$ possible outcomes, instead of needing to distinguish between the $2^N$ possible final states. We estimate the effect of measurement error for this detection scheme. 

Finally, we quantify the effect of atom loss. By using the capability of detecting whether an atom was lost, we can post-select only lossless outcomes, which will result in a reduced data-acquisition rate. If the total number of experiments (AI cycles) is fixed, a decrease in the number of lossless experiments can be interpreted as an effective deterioration of phase uncertainty per one experiment.

\section{Background}
\label{sec:background}

\begin{figure}[t]
	\centering
	\includegraphics[width=0.55\linewidth]{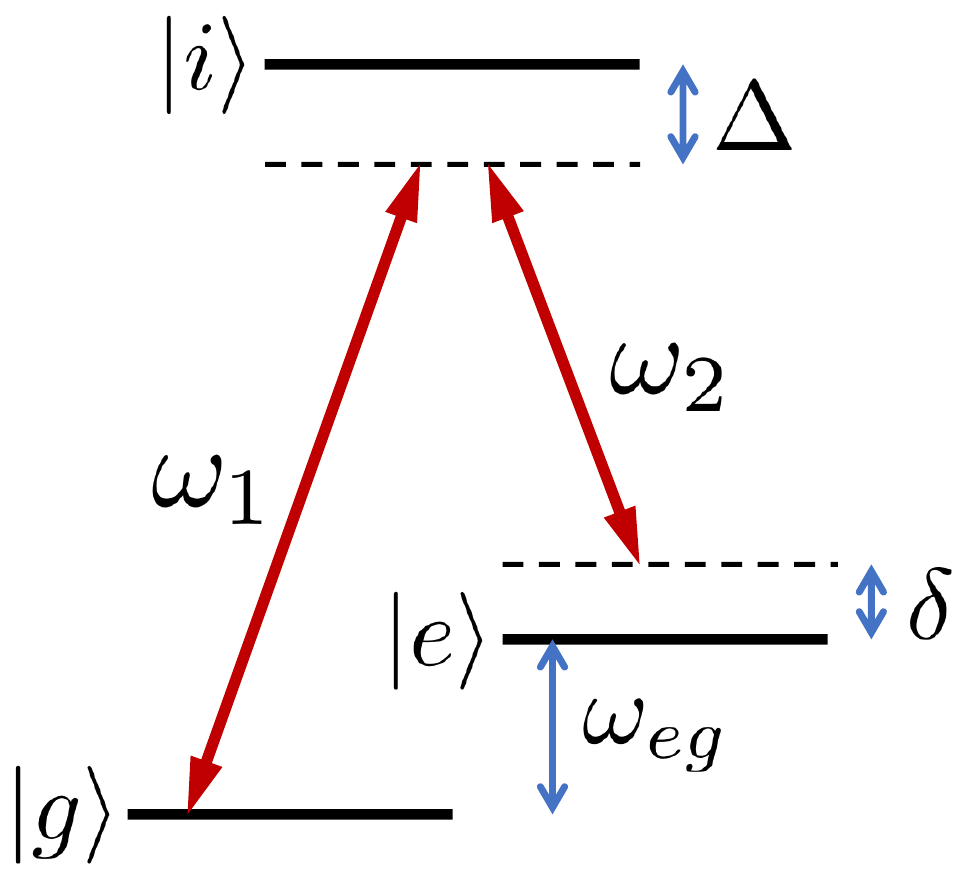}
	\caption{A scheme of a two-photon stimulated Raman transition in a three-level atom. Two light fields with frequencies $\omega_1$ and $\omega_2$ couple states $|g\rangle$ and $|e\rangle$ via the intermediate state $|i\rangle$. The one-photon and two-photon detunings are $\Delta = \omega_1 - (\omega_i - \omega_g)$ and $\delta = \omega_1 - \omega_2 - \omega_{eg}$, respectively.}
	\label{fig:Raman_scheme}
\end{figure}

Precision AI experiments~\cite{Kasevich.PRL.67.181.1991, Kasevich.APB.54.321.1992, Young.chapter.1997} typically employ the clock transition between the ground-state hyperfine levels with $m_F = 0$ (this transition is magnetic field insensitive). In particular, for $^{133}$Cs atoms, the electronic ground state is $6S_{1/2}$, and the clock transition is between the hyperfine levels $|g\rangle = |F = 3, m_F = 0\rangle$ and $|e\rangle = |F = 4, m_F = 0\rangle$. The standard experimental approach~\cite{Kasevich.PRL.67.181.1991, Kasevich.APB.54.321.1992} is based on driving the two-photon stimulated Raman transition between these levels, via an intermediate level $|i\rangle$, as schematically shown in Fig.~\ref{fig:Raman_scheme}. 
The model of the atom-field interaction during this transition is based on the Hamiltonian~\cite{Young.chapter.1997}:
\begin{equation}
\label{eq:Ham-Raman}
H = -\frac{\hbar^2}{2 m}\nabla^2 + \hbar \omega_g |g\rangle \langle g| + \hbar \omega_e |e\rangle \langle e| + \hbar \omega_i |i\rangle \langle i| 
- \mathbf{d} \cdot \mathbf{E} ,  
\end{equation}
where $m$ is the atom's mass, $\hbar \omega_g$, $\hbar \omega_e$, and $\hbar \omega_i$ are the energies of the three atomic levels, $\mathbf{d}$ is the atom's dipole operator, and $\mathbf{E}$ is the electric field of two Raman beams:
\begin{equation}
\mathbf{E} = \mathbf{E}_1 \cos(\mathbf{k}_1 \cdot \mathbf{x} - \omega_1 t + \phi_1) 
+ \mathbf{E}_2 \cos(\mathbf{k}_2 \cdot \mathbf{x} - \omega_2 t + \phi_2) .  
\end{equation}
Here, $\mathbf{x}$ is the atom's position, and each Raman field is characterized by its amplitude $\mathbf{E}_j$, wave vector $\mathbf{k}_j$, frequency $\omega_j$ and phase $\phi_j$ ($j =1,2$). 

A solution to the Schr\"{o}dinger equation governed by the Hamiltonian of Eq.~\eqref{eq:Ham-Raman} can be generally expressed as a superposition:
\begin{align}
|\psi(t)\rangle = \int d^3 p\, \sum_{\alpha} c_{\alpha,\mathbf{p}}(t) |\alpha,\mathbf{p}\rangle (t), \label{eq:psi-general} \\
|\alpha,\mathbf{p}\rangle (t) = e^{-i \left(\omega_{\alpha} + {\textstyle \frac{\mathbf{p}^2} {2m\hbar} } \right) t} |\alpha,\mathbf{p}\rangle ,
\label{eq:basis-FFE}
\end{align}
where the basis state $|\alpha,\mathbf{p}\rangle = |\alpha\rangle \otimes |\mathbf{p}\rangle$ corresponds to an atom in its internal state $|\alpha\rangle$ ($\alpha = \{g,e,i\}$) and in its momentum eigenstate $|\mathbf{p}\rangle$, whose position-space representation is $\psi_{\mathbf{p}}(\mathbf{x}) = e^{i \mathbf{p} \cdot \mathbf{x}/\hbar}$. The time dependence in Eq.~\eqref{eq:basis-FFE} captures the field-free evolution. 
Due to momentum conservation, for a given $\mathbf{p}$, the Hamiltonian of Eq.~\eqref{eq:Ham-Raman} only couples the states $|g,\mathbf{p}\rangle$, $|i,\mathbf{p} + \hbar \mathbf{k}_1 \rangle$, and $|e,\mathbf{p} + \hbar \mathbf{K} \rangle$, where $\mathbf{K} = \mathbf{k}_1 - \mathbf{k}_2$. The momentum kick $\hbar \mathbf{K}$ is maximized by using counter-propagating Raman beams, for which $\mathbf{K} \approx 2 \mathbf{k}_1$.

Adiabatic elimination of the intermediate level $|i\rangle$ results in a model for an effective two-level system, whose Hamiltonian couples the states $|g,\mathbf{p}\rangle$ and $|e,\mathbf{p} + \hbar \mathbf{K} \rangle$. In the field-free evolving basis $\{ |g,\mathbf{p}\rangle(t) , |e,\mathbf{p}+\hbar \mathbf{K}\rangle(t)\}$, this effective Hamiltonian is~\cite{Young.chapter.1997}
\begin{equation}
\label{eq:Ham-eff}
H_{\mathrm{eff}} = \hbar \begin{bmatrix}
\Omega_g^{\mathrm{AC}} & ( \Omega_{\mathrm{eff}}/2) e^{i(\delta_{12}t + \phi_{12} )} \\
( \Omega_{\mathrm{eff}}/2) e^{-i(\delta_{12}t + \phi_{12} )} & \Omega_e^{\mathrm{AC}}
\end{bmatrix} ,
\end{equation}
where
\begin{align}
& \Omega_g^{\mathrm{AC}} = \frac{|\Omega_g|^2}{4\Delta} , \quad 
\Omega_e^{\mathrm{AC}} = \frac{|\Omega_e|^2}{4\Delta} , \\
& \Omega_g = - \langle i | \mathbf{d} \cdot \mathbf{E}_1 |g\rangle / \hbar , \quad
\Omega_e = - \langle i | \mathbf{d} \cdot \mathbf{E}_2 |e\rangle / \hbar , \\
\label{eq:delta12}
& \delta_{12} = \omega_{12} - \left( \omega_{eg} + \frac{\mathbf{p} \cdot \mathbf{K}}{m} + \frac{\hbar |\mathbf{K}|^2}{2 m} \right) , \\
\label{eq:Omega-eff}
& \Omega_{\mathrm{eff}} = \frac{\Omega_e^{\ast} \Omega_g}{2 \Delta} e^{i \phi_{12}} , \quad
\phi_{12} = \phi_1 - \phi_2 .
\end{align}
Here, $\omega_{12} = \omega_1 - \omega_2$ is the frequency difference between the two Raman fields, $\omega_{eg} = \omega_e - \omega_g$ is the frequency difference between the hyperfine levels, and $\Delta = \omega_1 - (\omega_i - \omega_g)$ is the detuning from the optical resonance, as shown in Fig.~\ref{fig:Raman_scheme}. The diagonal elements of the Hamiltonian in Eq.~\eqref{eq:Ham-eff} are the ac Stark shifts of levels $|g\rangle$
and $|e\rangle$, and the relative ac Stark shift of the two levels is $\delta^{\mathrm{AC}} \equiv \Omega_e^{\mathrm{AC}} - \Omega_g^{\mathrm{AC}}$. The effective detuning from the Raman resonance, $\delta_{12}$ in Eq.~\eqref{eq:delta12}, includes the two-photon detuning, $\delta = \omega_{12} - \omega_{eg}$, the detuning due to the Doppler shift, $-\mathbf{p} \cdot \mathbf{K} / m$, and the detuning due to the atom's recoil energy, $-\hbar |\mathbf{K}|^2 / 2 m$. The phase $\phi_{12}$ is chosen to make the effective Rabi frequency $\Omega_{\mathrm{eff}}$ a positive real number.

The Schr\"{o}dinger equation governed by the Hamiltonian of Eq.~\eqref{eq:Ham-eff} can be solved analytically to obtain the evolution operator for a single atom~\cite{Young.chapter.1997}. If the Raman fields are turned on at time $t$ and act for duration $\tau$, the resulting evolution operator, denoted as $U_t (\tau)$, is represented in the basis $\{ |g,\mathbf{p}\rangle(t) , |e,\mathbf{p}+\hbar \mathbf{K}\rangle(t)\}$ as
\begin{widetext}
\begin{equation}
U_t (\tau) = \begin{bmatrix}
e^{i\delta_{12}\tau/2} \left[ \cos\left( \frac{\Omega_r \tau}{2} \right) + i \cos\Theta \sin\left( \frac{\Omega_r \tau}{2} \right) \right] & 
-i e^{i\delta_{12}\tau/2} e^{i(\phi_t+\phi_{12})} \sin\Theta \sin\left( \frac{\Omega_r \tau}{2} \right) \\[1em]
-i e^{-i\delta_{12}\tau/2} e^{-i(\phi_t+\phi_{12})} \sin\Theta \sin\left( \frac{\Omega_r \tau}{2} \right) & 
e^{-i\delta_{12}\tau/2} \left[ \cos\left( \frac{\Omega_r \tau}{2} \right) - i \cos\Theta \sin\left( \frac{\Omega_r \tau}{2} \right) \right] 
\end{bmatrix} ,
\label{eq:U-general}
\end{equation}
\end{widetext}
where we omitted a global phase factor $e^{-i (\Omega_g^{\mathrm{AC}} + \Omega_e^{\mathrm{AC}}) \tau/2}$ and defined
\begin{gather*}
\Omega_r \equiv \sqrt{\Omega_\text{eff}^2 + (\delta_{12} - \delta^{\mathrm{AC}})^2} ,  \\
\sin\Theta \equiv \Omega_\text{eff}/\Omega_r, \quad \cos\Theta \equiv - (\delta_{12} - \delta^{\mathrm{AC}})/\Omega_r , 
\end{gather*}
with $0 \leq \Theta \leq \pi$, and
\begin{equation}
\phi_t \equiv \int_0^t \delta_{12}(t') d t' .
\label{eq:phi_t}
\end{equation}
If the atom moves at a constant velocity, then $\delta_{12}$ is constant, and $\phi_t = \delta_{12} t$. However, if the atom accelerates, $\delta_{12}$ is time dependent, and Eq.~\eqref{eq:phi_t} should be used. Note that Eq.~\eqref{eq:U-general} neglects the time dependence of $\delta_{12}$ during the pulse, assuming that the pulse duration $\tau$ is very short compared to periods of free evolution. Since the time dependence for the field-free evolution is included in Eq.~\eqref{eq:basis-FFE}, the evolution operator for a period of field-free evolution is represented in the basis $\{ |g,\mathbf{p}\rangle(t) , |e,\mathbf{p}+\hbar \mathbf{K}\rangle(t)\}$ by the two-by-two identity matrix. 

We will use the general form~\eqref{eq:U-general} in Sec.~\ref{sec:Doppler-shift} where we explicitly take into account the effect of the initial momentum spread of the atoms, which leads to a detuning uncertainty via the Doppler shift term in Eq.~\eqref{eq:delta12}. However, in Secs.~\ref{sec:ideal}--\ref{sec:phase-noise} we assume that the Raman detuning is much smaller than the effective Rabi frequency, $|\delta_{12}| \ll \Omega_\text{eff}$. This assumption is satisfied if three conditions are met: (1) short, intense driving fields are used to make $\Omega_\text{eff}$ large, (2) the momentum distribution of the atoms is sufficiently narrow to make $|\delta_{12}|$ small for all relevant momentum components, and (3) laser frequency chirping is used to compensate for the evolving Doppler shift due to the acceleration of the atom (see \cite{Kasevich.APB.54.321.1992, Young.chapter.1997} for more details). Under this assumption, we neglect terms on the order of $|\delta_{12} - \delta^{\mathrm{AC}}| / \Omega_\text{eff}$ and $|\delta_{12}| \tau$ in Eq.~\eqref{eq:U-general} to obtain:
\begin{equation}
U_t (A) = \begin{bmatrix}
\cos\frac{A}{2} & -i e^{i(\phi_t+\phi_{12})} \sin\frac{A}{2} \\
-i e^{-i(\phi_t+\phi_{12})} \sin\frac{A}{2} & \cos\frac{A}{2}
\end{bmatrix} ,
\label{eq:U-1}
\end{equation}
where $A \equiv \Omega_{\mathrm{eff}} \tau$ is the pulse area and, since $\tau$ enters only via $A$, we changed the notation from $U_t (\tau)$ to $U_t (A)$. We will use this approximation for the pulse evolution operator throughout this paper, except for Sec.~\ref{sec:Doppler-shift} and Appendix~\ref{sec:app-corrections}. 

Since atoms interact independently with light fields, the evolution operator for $N$ atoms is 
\begin{equation}
\label{eq:U}
U_t (A) = \bigotimes_{k=1}^N U_t^{(k)}(A) ,
\end{equation}
where $U_t^{(k)}(A)$ is the evolution operator for the $k$th atom.

A typical AI operation includes three pulses: $\pi/2$--$\pi$--$\pi/2$, with two periods of field-free evolution, each of duration $T$, between the pulses. The evolution operator for one entire cycle of the AI operation is
\begin{equation}
U_{\text{tot}} = U_{t_3} (\pi/2) U_{t_2} (\pi) U_{t_1} (\pi/2) ,
\end{equation}
where $t_1$ is the starting time of the first $\pi/2$ pulse and, correspondingly, $t_2 = t_1 + \tau/2 + T$ and $t_3 = t_1 + 3 \tau/2 + 2 T$, assuming that the $\pi$ pulse has a duration $\tau$ and each of the $\pi/2$ pulses has a duration $\tau/2$. In what follows, without loss of generality, we set $t_1 = 0$. We also assume that the duration of each pulse (typically, $\tau \sim 1~\mu\mathrm{s}$) is negligible compared to the duration of the free evolution (typically, $T \sim 1~\mathrm{ms}$). Under this assumption, we neglect terms on the order of $|\delta_{12}| \tau$ [which is consistent with the approximation we made in deriving Eq.~\eqref{eq:U-1}] and obtain:
\begin{equation}
U_{\text{tot}} = U_{2T} (\pi/2) U_{T} (\pi) U_{0} (\pi/2) .
\label{eq:U-tot-3-pulses-operator-form}
\end{equation}

Immediately after the final pulse, a state-dependent detection of the atoms is used to measure the interference. Specifically, we assume that one measures the expectation value of the parity operator,
\begin{equation}
\label{eq:Parity-operator}
\Pi = \bigotimes_{k=1}^{N} \sigma_z^{(k)} ,
\end{equation} 
where $\sigma_z^{(k)} = |g\rangle_k\, {}_k \langle g | - |e\rangle_k\, {}_k \langle e |$ is the Pauli $z$ matrix for the $k$th atom's spin. Since the parity operator (or any operator corresponding to a measurement of the populations of the atomic levels and hence diagonal in the basis $\{|g\rangle_k, |e\rangle_k\}$) is invariant under the field-free evolution, its expectation value at the final time $t = 2 T$ is given by
\begin{equation}
\label{eq:Parity-EV-general}
\braket{\Pi} = \langle \Psi(0) | U_{\text{tot}}^{\dagger} (\Pi \otimes \openone_{\mathbf{p}}) U_{\text{tot}} |\Psi(0)\rangle ,
\end{equation} 
where $\openone_{\mathbf{p}} = \bigotimes_{k=1}^{N} \int d^3 p_k | \mathbf{p}_k \rangle \langle \mathbf{p}_k |$ is the identity operator for the motional degrees of freedom and $|\Psi(0)\rangle$ is the initial state of the system of $N$ atoms. 

\section{The ideal case} 
\label{sec:ideal}

As a reference point, we first consider the case of an AI without errors. In this ideal case, we ignore all physical errors in the Raman pulses, which means that both $A$ and $\phi_{12}$ in the evolution operator of Eq.~\eqref{eq:U-1} are precisely known.

The evolution operator $U_t (A)$ of Eq.~\eqref{eq:U-1} represents the rotation by an angle $A$ about the axis $\hat{\mathbf{x}} \cos(\phi_t+\phi_{12}) - \hat{\mathbf{y}} \sin(\phi_t+\phi_{12})$. The parity expectation value is independent of the specific choice of $\phi_{12}$, so, for the sake of simplicity, we use $\phi_{12} = -\pi/2$ throughout the rest of this paper. With this choice, the evolution operator of Eq.~\eqref{eq:U-1} becomes
\begin{equation}
U_t (A) = \begin{bmatrix}
\cos\frac{A}{2} & - e^{i\phi_t} \sin\frac{A}{2} \\
e^{-i\phi_t} \sin\frac{A}{2} & \cos\frac{A}{2} \end{bmatrix} .
\label{eq:y-rotation}
\end{equation}
Specifically, the evolution operators for the three AI pulses are
\begin{align}
& U_{0} (\pi/2) = \frac{1}{\sqrt{2}} \begin{bmatrix} 1 & -1 \\ 1 & 1 \end{bmatrix} ,  \label{eq:pi-over-2-first} \\
& U_{T} (\pi) = \begin{bmatrix} 0 & -e^{i\phi_{T}} \\ e^{-i\phi_{T}} & 0 \end{bmatrix} ,  \label{eq:pi-second} \\
& U_{2T} (\pi/2) = \frac{1}{\sqrt{2}} \begin{bmatrix} 1 & -e^{i\phi_{2T}} \\ e^{-i\phi_{2T}} & 1 \end{bmatrix} .  \label{eq:pi-over-2-third}
\end{align}
Note that in Eq.~\eqref{eq:pi-over-2-first} we used the fact that $\phi_{t=0} = 0$.

\begin{figure}[htbp]
	\centering
	\includegraphics[width=0.8\linewidth]{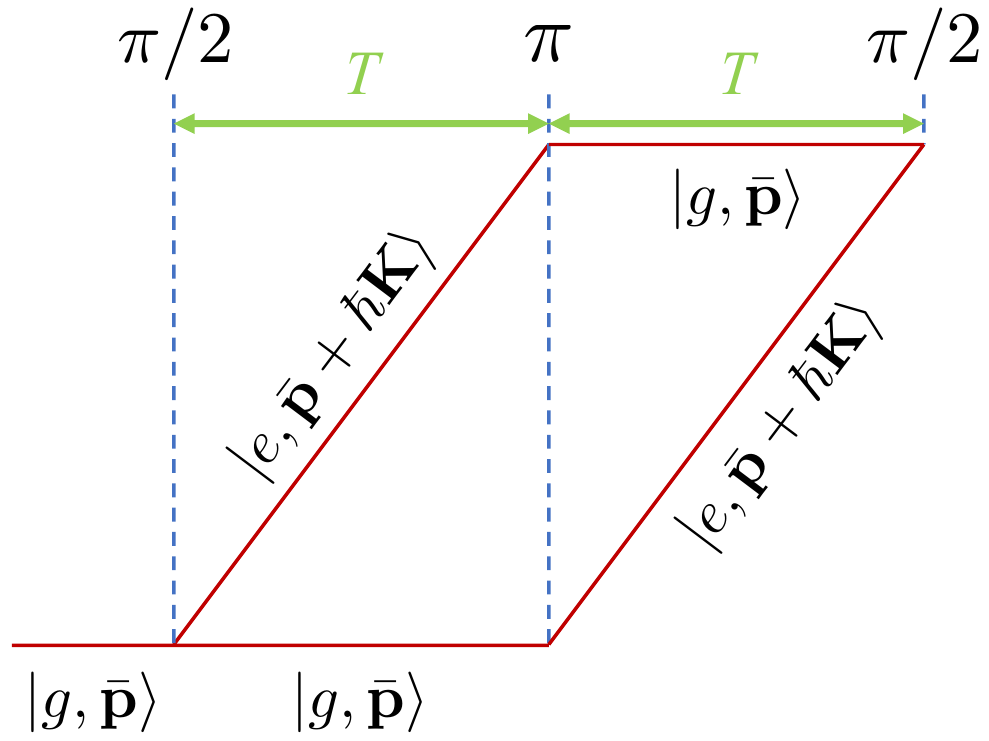}
	\caption{A scheme of an AI with the $\pi/2$--$\pi$--$\pi/2$ sequence of pulses acting on the initial state $|g,\bar{\mathbf{p}}\rangle$. Trajectories of the wavefunction components that differ by the momentum $\hbar \mathbf{K}$ are indicated.}
	\label{fig:AI_scheme_3_pulses}
\end{figure}

\subsection{Ideal interferometer with independent atoms}
\label{sec:ideal-independent}

For an AI with $N$ independent atoms, each atom interferes with itself \cite{Parazzoli.PRL.109.230401.2012}. Let us first consider one atom which is initially in the state 
\begin{equation}
|\psi\rangle_{\text{in}} \equiv |\psi(0)\rangle = |g,\bar{\mathbf{p}}\rangle = |g\rangle \otimes |\bar{\mathbf{p}}\rangle ,
\end{equation}
where 
\begin{equation}
|\bar{\mathbf{p}}\rangle = \int d^3 p\, \tilde{\psi}_{\bar{\mathbf{p}}}(\mathbf{p}) | \mathbf{p} \rangle 
\end{equation}
denotes a motional state of the atom with the average momentum $\bar{\mathbf{p}} = \int d^3 p\, |\tilde{\psi}_{\bar{\mathbf{p}}}(\mathbf{p})|^2\, \mathbf{p}$. The AI operation with the pulse sequence $\pi/2$--$\pi$--$\pi/2$ acting on the initial state $|g,\bar{\mathbf{p}}\rangle$ is shown schematically in Fig.~\ref{fig:AI_scheme_3_pulses}. Since this initial state is a linear superposition of the basis states $|g,\mathbf{p}\rangle$, we first calculate the evolution for a basis state with a given $\mathbf{p}$ and integrate over the momentum distribution only at the final step when computing the parity expectation value via Eq.~\eqref{eq:Parity-EV-general}.

Using Eqs.~\eqref{eq:pi-over-2-first}--\eqref{eq:pi-over-2-third}, we find that the evolution operator $U_{\text{tot}}$ of Eq.~\eqref{eq:U-tot-3-pulses-operator-form} has the following matrix form:
\begin{equation}
U_{\text{tot}} = \frac{1}{2} \begin{bmatrix}
- e^{i \phi_T} (1+ e^{i\phi}) & - e^{i \phi_T} (1 - e^{i\phi}) \\
e^{-i \phi_T} (1 - e^{-i\phi})& - e^{-i \phi_T} (1 + e^{-i\phi})
\end{bmatrix} \label{eq:U-tot-3-pulses} 
\end{equation}
and
\begin{align}
U_{\text{tot}}  |g,\mathbf{p}\rangle =\, & {\textstyle \frac{1}{2}} 
\big[ e^{-i\phi_{T}} \big( 1 - e^{-i \phi} \big) |e,\mathbf{p}+\hbar \mathbf{K}\rangle \nonumber \\
& - e^{i\phi_{T}} \big( 1 + e^{i \phi} \big)  |g,\mathbf{p}\rangle \big] ,
\label{eq:AI-output}
\end{align}
where 
\begin{equation}
\phi \equiv \phi_{2T} - 2 \phi_{T} .
\label{eq:phi-1}
\end{equation}
By noting that
\begin{equation}
\phi = \int_T^{2T} \delta_{12} (t') dt' - \int_0^T \delta_{12} (t') dt' ,
\label{eq:phi-2}
\end{equation}
we see that $\phi$ is the difference between the phases accumulated via the Raman detuning in the second and first halves of the atom's trajectory. The phase $\phi$ is zero if the atom moves at a constant velocity. If the atom moves with a constant acceleration $\mathbf{a}$, then $\delta_{12}$ depends linearly on time. In the laboratory frame, the atom's momentum changes with time as $\mathbf{p}(t) = \mathbf{p}(0) + m \mathbf{a} t$. In the frame that accelerates with the atom, the momentum $\mathbf{p}$ is constant, but the Raman frequency experiences a time-dependent Doppler shift: $\omega_{12}(t) = \omega_{12}(0) - \mathbf{K} \cdot \mathbf{a} t$. Regardless of which frame is used, the time-dependent part of the Raman detuning is $- \mathbf{K} \cdot \mathbf{a} t$, which yields
\begin{equation}
\phi = - \mathbf{K} \cdot \mathbf{a} T^2 .
\label{eq:phi-3}
\end{equation}
Since $\phi$ is independent of $\mathbf{p}$, we do not need to know the specific momentum distribution $|\tilde{\psi}_{\bar{\mathbf{p}}}(\mathbf{p})|^2$ of the initial state $|\psi(0)\rangle$ to compute the expectation value of the parity operator via Eq.~\eqref{eq:Parity-EV-general}. Using Eqs.~\eqref{eq:Parity-EV-general}~and~\eqref{eq:AI-output}, we obtain:
\begin{equation}
\braket{\Pi} = \frac{1}{4} \left( \left| 1 + e^{i \phi} \right|^2 - \left| 1 - e^{-i \phi} \right|^2 \right) = \cos\phi .
\label{eq:Pi-expect-0}
\end{equation}
This result holds for any momentum distribution of the initial state as long as it is sufficiently narrow for the approximation used for Eq.~\eqref{eq:U-1} to be valid.

For the AI with $N$ independent atoms, the initial state is
\begin{equation}
|\Psi\rangle_{\text{in}} = \bigotimes_{k=1}^{N}  |g,\bar{\mathbf{p}}_k \rangle_k 
= \bigotimes_{k=1}^{N}  \int d^3 p_k\, \tilde{\psi}_{\bar{\mathbf{p}}_k}(\mathbf{p}_k)  |g,\mathbf{p}_k \rangle_k .
\label{eq:AI-state-in}
\end{equation}
Note that, for the sake of generality, we assume that the wavefunctions $\tilde{\psi}_{\bar{\mathbf{p}}_k}$ can be different for different $k$, although, as we will see below, this assumption does not affect the result for the parity expectation value.
The evolution operator for one atom is given by Eq.~\eqref{eq:U-tot-3-pulses} or, explicitly for the $k$th atom, by Eq.~\eqref{eq:app:U-tot-3-pulses}, where
\begin{equation}
\phi_k \equiv \phi_{2T}^{(k)} - 2 \phi_{T}^{(k)} . 
\label{eq:phi-k}
\end{equation}
For the $k$th atom, the phases $\phi_{T}^{(k)}$ and $\phi_{2T}^{(k)}$ depend on the momentum $\mathbf{p}_k$ through the Doppler term $-\mathbf{p}_k \cdot \mathbf{K} / m$ in the Raman detuning $\delta_{12}^{(k)}$. However, if the $k$th atom moves with a constant acceleration $\mathbf{a}_k$, then, analogously to Eq.~\eqref{eq:phi-3}, $\phi_k = -\mathbf{K} \cdot \mathbf{a}_k T^2$ is independent of $\mathbf{p}_k$. Therefore, we once again do not need to know the specific momentum distributions $|\tilde{\psi}_{\bar{\mathbf{p}}_k}(\mathbf{p}_k)|^2$ to compute the expectation value of the parity operator via Eq.~\eqref{eq:Parity-EV-general}. It also does not matter whether $\tilde{\psi}_{\bar{\mathbf{p}}_k}$ is the same for all $k$ or not. The derivation of the parity expectation value is described in detail in Appendix~\ref{sec:app-derivation}. Specifically, for the initial state of Eq.~\eqref{eq:AI-state-in}, we obtain:
\begin{equation}
\braket{\Pi} = \prod_{k=1}^{N} \cos\phi_k .
\end{equation}
If all atoms experience the same constant acceleration $\mathbf{a}_k = \mathbf{a}$, then $\phi_k = \phi = - \mathbf{K} \cdot \mathbf{a} T^2$, and we find:
\begin{equation}
\label{eq:parity-EV}
\braket{\Pi} = \cos^N\! \phi .
\end{equation}

The uncertainty of the measured $\phi$ value is
\begin{equation}
\label{Delta-phi}
\Delta\phi = \frac{\Delta\Pi}{\left| \partial \braket{\Pi} /  \partial \phi \right| },
\end{equation}
where $\Delta\Pi = \sqrt{ \langle \Pi^2 \rangle - \braket{\Pi}^2} = \sqrt{1 - \braket{\Pi}^2}$. Using Eq.~\eqref{eq:parity-EV}, we obtain:
\begin{equation}
\label{Delta-phi-1}
\Delta\phi = \frac{ \sqrt{1 - \cos^{2N} \phi}}{N \left| \cos^{N-1} \phi \, \sin\phi \right| }.
\end{equation}
It is easy to verify that the phase uncertainty of Eq.~\eqref{Delta-phi-1} is minimized for $\phi = n \pi$, where $n = 0, \pm 1, \pm 2, \ldots$. Then we obtain:
\begin{equation}
\label{Delta-phi-2}
\Delta\phi = \frac{1}{\sqrt{N}} .
\end{equation}
As expected for the AI with independent atoms, this phase uncertainty scales according to the SQL.

\subsection{Ideal interferometer with entangled atoms}
\label{sec:ideal-entangled}

In order to surpass the SQL, we propose an AI using an entangled state of atoms. Specifically, we consider a collection of $N$ atoms in the GHZ state \cite{GHZ.arxiv.0712.0921}:
\begin{align}
|\Psi\rangle_{\mathrm{in}} & = | \mathrm{GHZ}_N \rangle \nonumber \\
& = \frac{1}{\sqrt{2}} \left( 
\bigotimes_{k=1}^{N}  |g,\bar{\mathbf{p}}_k \rangle_k +
\bigotimes_{k=1}^{N}  |e,\bar{\mathbf{p}}_k+\hbar \mathbf{K}\rangle_k \right) ,
\label{eq:GHZ-state-in} 
\end{align}
where
\begin{equation}
|\alpha,\bar{\mathbf{p}}_k+\hbar \mathbf{K}_{\alpha}\rangle_k 
= \int d^3 p_k\, \tilde{\psi}_{\bar{\mathbf{p}}_k}(\mathbf{p}_k)  |\alpha,\mathbf{p}_k+\hbar \mathbf{K}_{\alpha}\rangle_k ,
\end{equation}
for $\alpha = \{g,e\}$ with $\mathbf{K}_{g} = 0$ and $\mathbf{K}_{e} = \mathbf{K}$.

\begin{figure}[t]
	\centering
	\includegraphics[width=0.9\linewidth]{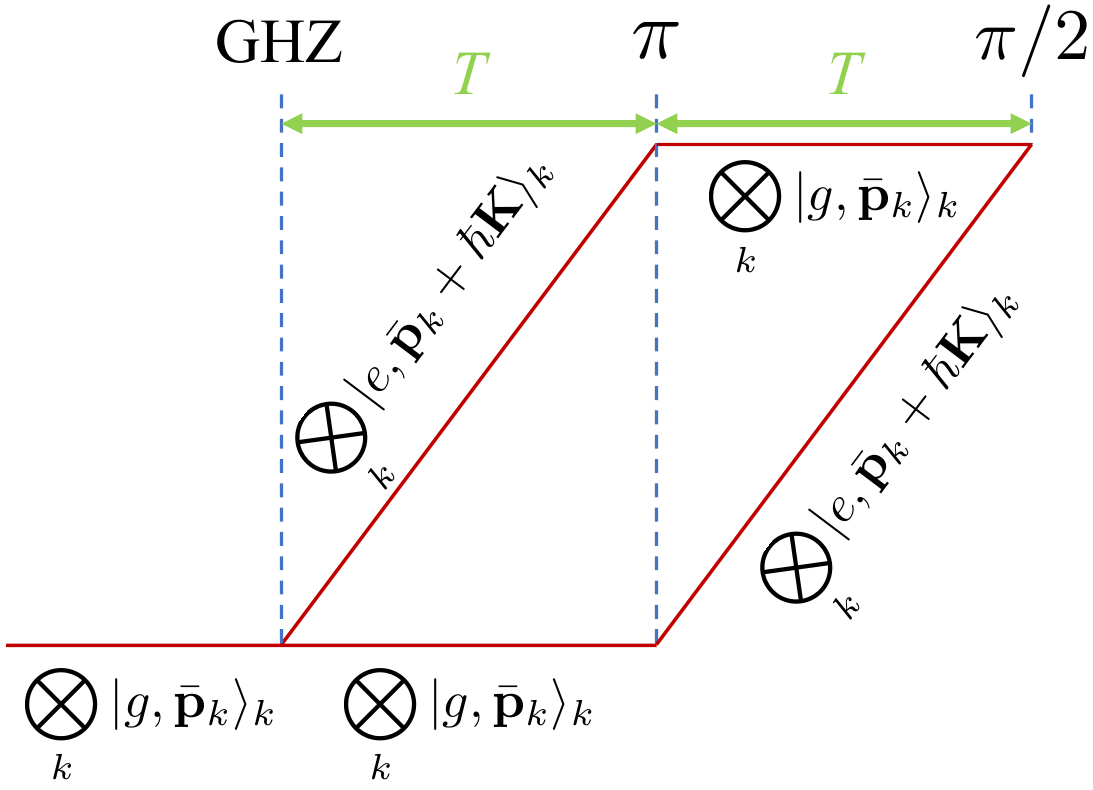}
	\caption{A scheme of an entanglement-enhanced AI, where the generation of the GHZ state replaces the first $\pi/2$ pulse. Trajectories of the wavefunction components that differ by the momentum $\hbar \mathbf{K}$ are indicated.}
	\label{fig:AI_scheme_2_pulses}
\end{figure}

Normally, the term ``GHZ state'' is used only for $N \geq 3$, however, the state of Eq.~\eqref{eq:GHZ-state-in} is defined for any integer $N \geq 1$. For $N = 1$, this is the single-atom state $U_{0} (\pi/2) |g,\bar{\mathbf{p}}\rangle = \frac{1}{\sqrt{2}} ( |g,\bar{\mathbf{p}}\rangle + |e,\bar{\mathbf{p}} + \hbar \mathbf{K}\rangle )$. For $N = 2$, this is the two-atom Bell state $|\Phi^{+}\rangle$ \cite{Nielsen.Chuang.2010}.

As shown in Fig.~\ref{fig:AI_scheme_2_pulses}, the proposed entanglement-enhanced AI scheme differs from the standard operation described in Sec.~\ref{sec:ideal-independent} above by the replacement of the first $\pi/2$ pulse at time $t = 0$ by a process that generates the GHZ input state of Eq.~\eqref{eq:GHZ-state-in}. The rest of the AI operation, including the $\pi$ pulse at time $t = T$ and the $\pi/2$ pulse at time $t = 2 T$, remains the same. 
Correspondingly, $U_{\text{tot}}^{(k)} = U_{2T}^{(k)} (\pi/2) U_{T}^{(k)} (\pi)$, and, using Eqs.~\eqref{eq:pi-second} and \eqref{eq:pi-over-2-third}, we obtain:
\begin{equation}
U_{\text{tot}}^{(k)} = \frac{1}{\sqrt{2}} \begin{bmatrix}
- e^{-i \phi_T^{(k)}} e^{i \phi_{2T}^{(k)}} & - e^{i \phi_T^{(k)}} \\
e^{-i \phi_T^{(k)}} & - e^{i \phi_T^{(k)}} e^{-i \phi_{2T}^{(k)}}
\end{bmatrix} . 
\label{eq:U-tot-2-pulses} 
\end{equation}

The derivation of the parity expectation value is described in detail in Appendix~\ref{sec:app-derivation}. Specifically, for the initial GHZ state of Eq.~\eqref{eq:GHZ-state-in} and the evolution operator of Eq.~\eqref{eq:U-tot-2-pulses}, we obtain:
\begin{equation}
\braket{\Pi} = \cos\left( \sum_{k=1}^N \phi_k \right) ,
\label{parity-EV-GHZ-k}
\end{equation}
where $\phi_k$ is given by Eq.~\eqref{eq:phi-k}. Since $\phi_k$ is independent of $\bar{\mathbf{p}}_k$, the result of Eq.~\eqref{parity-EV-GHZ-k} holds for any momentum distribution of the initial state as long as it is sufficiently narrow for the approximation of Eq.~\eqref{eq:U-1} to be valid.

If all atoms experience the same constant acceleration $\mathbf{a}$, then $\phi_k = \phi = \mathbf{K} \cdot \mathbf{a} T^2$, and we obtain:
\begin{equation}
\label{eq:parity-EV-GHZ}
\braket{\Pi} = \cos(N \phi) .
\end{equation}
The phase uncertainty is obtained by substituting Eq.~\eqref{eq:parity-EV-GHZ} into Eq.~\eqref{Delta-phi}, which yields, for any value of $\phi$,
\begin{equation}
\label{Delta-phi-GHZ}
\Delta\phi = \frac{1}{N} .
\end{equation}
This is the HL, the ultimate scaling of the phase uncertainty with respect to the number of particles allowed by the laws of quantum mechanics.

\section{Effect of imperfect preparation of the initial entangled state}
\label{sec:imperfect-state-prep}

In practice, inevitable noise and imperfections can preclude the achievement of the HL. Our goal is to quantify the deviation from the HL due to these practically relevant errors.

First, we investigate the effect of imperfections in the GHZ state at the input to the AI. We consider two types of state imperfections: a random relative phase between the two components of the GHZ state and the admixture of a noise state.\footnote{We do not consider here a state imperfection associated with a case where some of the atoms are unentangled but still contribute to the signal. In experiments, in which high-fidelity entangled states of atomic spins are generated via Rydberg-mediated interactions~\cite{Jau.NatPhys.12.71.2016, Martin.SAND-report.2018, Levine.PRL.121.123603.2018, Graham.PRL.123.230501.2019, Levine.PRL.123.170503.2019, Omran.Science.365.570.2019}, the atoms are individually controlled. Therefore, in a well-designed experiment, a situation, in which a portion of the atoms would remain unentangled and contaminate the signal from the GHZ state, is unrealistic.} This input state can be represented by the density matrix:
\begin{align}
& \rho_{\mathrm{in}} = (1-q_{\zeta}) | \Psi(\beta) \rangle \langle \Psi(\beta) | + q_{\zeta} | \zeta \rangle \langle \zeta | , \label{eq:GHZ-state-imperfect-in} \\
& | \Psi(\beta) \rangle = \frac{1}{\sqrt{2}} \left( 
\bigotimes_{k=1}^{N}  |g,\bar{\mathbf{p}}_k \rangle_k + e^{i\beta}
\bigotimes_{k=1}^{N}  |e,\bar{\mathbf{p}}_k+\hbar \mathbf{K}\rangle_k \right) , \label{eq:Psi-beta-state} \\
& | \zeta \rangle = \bigotimes_{k=1}^{N} \left( 
\cos \frac{\vartheta_k}{2} |g,\bar{\mathbf{p}}_k \rangle_k + e^{i\varphi_k} \sin \frac{\vartheta_k}{2} |e,\bar{\mathbf{p}}_k+\hbar \mathbf{K}\rangle_k \right) .
\label{eq:zeta-state}
\end{align}
Here, $| \Psi(\beta) \rangle$ is the GHZ state with a random relative phase $\beta$ between the two components, and $| \zeta \rangle$ is a noise state admixed with probability $q_{\zeta}$ ($0 \leq q_{\zeta} \leq 1$). With the initial state of Eq.~\eqref{eq:GHZ-state-imperfect-in} and the same AI operation as described in Sec.~\ref{sec:ideal-entangled}, i.e., the evolution operator $U_{\text{tot}}^{(k)}$ of Eq.~\eqref{eq:U-tot-2-pulses}, the expectation value of the parity operator is (see Appendix~\ref{sec:app-derivation} for details of the derivation)
\begin{align}
\braket{\Pi} =\,  & (1-q_{\zeta}) \cos\left( \sum_{k=1}^N \phi_k - \beta \right) \nonumber \\
& + q_{\zeta} \prod_{k=1}^{N} \sin \vartheta_k \cos(\phi_k - \varphi_k) ,
\label{eq:Parity-imperfect-1}
\end{align}
where $\phi_k$ is given by Eq.~\eqref{eq:phi-k}. If we assume, as we have done previously, that all atoms experience the same constant acceleration $\mathbf{a}_k = \mathbf{a}$, then $\phi_k = \phi = \mathbf{K} \cdot \mathbf{a} T^2$, and $\sum_{k=1}^N \phi_k = N \phi$ in Eq.~\eqref{eq:Parity-imperfect-1}. We also assume that $\{\vartheta_k , \varphi_k \}$ and $\beta$ are independent random phase variables. Therefore, we are interested in the expectation value of the parity operator averaged over these random variables, $\overline{\braket{\Pi}}$. If $| \zeta \rangle$ represents a contribution from completely random noise, each $\vartheta_k$ has a uniform distribution with the probability density function $P(\vartheta_k) = 1/\pi$ on $[0,\pi]$, and each $\varphi_k$ has a uniform  distribution with $P(\varphi_k) = 1/(2\pi)$ on $[0,2\pi]$. Since 
$\int_0^{2\pi} \cos(\phi - \varphi_k) d \varphi_k = 0$ for any value of $\phi$, the contribution from the second term in Eq.~\eqref{eq:Parity-imperfect-1} is zero. With these assumptions, we obtain:
\begin{equation}
\overline{\braket{\Pi}} = \Pi_0 \cos( N \phi ) ,
\label{eq:Parity-imperfect-avg}
\end{equation}
and the amplitude of parity oscillations, $\Pi_0$, is given by
\begin{equation}
\Pi_0 = (1-q_{\zeta}) \overline{\cos\beta}
= (1-q_{\zeta}) \int_{-\pi}^{\pi} P(\beta) \cos\beta d \beta ,
\label{eq:Parity-imperfect-2}
\end{equation}
where $P(\beta)$ is the probability density function of a distribution for the random relative phase $\beta$. We assume that $\beta$ has the wrapped normal distribution~\cite{Mardia.book} with zero mean and variance $\sigma_{\beta}^2$, whose probability density function is given by
\begin{align}
P(\beta) =  f_{\mathrm{WN}}(\beta;0,\sigma_{\beta}^2) 
&=  \frac{1}{\sqrt{2\pi} \sigma_{\beta}} \sum_{j = -\infty}^{\infty} e^{-(\beta + 2\pi j)^2 / 2 \sigma_{\beta}^2} \nonumber \\
&= \frac{1}{2\pi} \sum_{n = -\infty}^{\infty} e^{i n \beta - n^2 \sigma_{\beta}^2 /2} .
\label{eq:WN-distribution}
\end{align}
Consequently, we obtain:
\begin{equation}
\Pi_0 = (1-q_{\zeta}) e^{-\sigma_{\beta}^2 /2} .
\label{eq:Parity-imperfect-2a}
\end{equation}
If the variance of the relative phase is small, $\sigma_{\beta}^2 \ll 1$, then Eq.~\eqref{eq:Parity-imperfect-2a} can be approximated as $\Pi_0 \approx (1-q_{\zeta}) (1-\sigma_{\beta}^2 /2)$, and if both errors are small, i.e., $\sigma_{\beta}^2 \ll 1$ and $q_{\zeta} \ll 1$, then
\begin{equation}
\Pi_0 \approx 1- (q_{\zeta} +\sigma_{\beta}^2 /2).
\label{eq:Parity-imperfect-3}
\end{equation}
A set of numerical simulations we performed for $N = 2$ confirms the accuracy of the approximate result in Eq.~\eqref{eq:Parity-imperfect-3}.

The fidelity of the initial state is 
\begin{equation}
\mathcal{F} = \langle \mathrm{GHZ}_N | \rho_{\mathrm{in}} | \mathrm{GHZ}_N \rangle .
\label{eq:fidelity-1}
\end{equation}
For the initial state of Eq.~\eqref{eq:GHZ-state-imperfect-in} and using the same averaging over the random variables as in the derivation of $\overline{\braket{\Pi}}$ above, we obtain the average fidelity:
\begin{equation}
\overline{\mathcal{F}} = ( 1 + \Pi_0 )/2 .
\label{eq:fidelity-2}
\end{equation}
Accordingly, the amplitude of parity oscillations can be expressed in terms of the initial state fidelity: $\Pi_0 = 2 \overline{\mathcal{F}} - 1$ or $\Pi_0 = 1 - 2\epsilon_{\text{prep}}$, where $\epsilon_{\text{prep}} \equiv 1 - \overline{\mathcal{F}}$ is the initial state preparation error. It follows from Eq.~\eqref{eq:Parity-imperfect-3} that $\epsilon_{\text{prep}} \approx \frac{1}{2} (q_{\zeta} +\sigma_{\beta}^2 /2)$ when the errors are small.

The phase uncertainty is obtained by substituting Eq.~\eqref{eq:Parity-imperfect-avg} into Eq.~\eqref{Delta-phi}, which yields
\begin{equation}
\label{Delta-phi-imperfect-1}
\Delta\phi = \frac{\sqrt{1 + (1 - \Pi_0^2) \cot^2(N \phi)}}{\Pi_0 N} .
\end{equation}
This uncertainty is minimized for the measurement at a dark fringe, where $\cos( N \phi ) = 0$. Then we obtain:
\begin{equation}
\label{Delta-phi-imperfect-2}
\Delta\phi = \frac{1}{\Pi_0 N} .
\end{equation}
In terms of the initial state preparation error, the phase uncertainty can be expressed as
\begin{equation}
\label{Delta-phi-imperfect-3}
\Delta\phi = \frac{1}{(1 - 2\epsilon_{\text{prep}}) N} \approx \frac{1}{[1- (q_{\zeta} + \sigma_{\beta}^2 /2)] N} ,
\end{equation}
where the approximate result is valid when the errors $q_{\zeta}$ and $\sigma_{\beta}^2$ are small.

Recent experiments reported the fidelity of the generated two-atom Bell state ranging from 0.89 to 0.97 \cite{Martin.SAND-report.2018, Levine.PRL.121.123603.2018, Graham.PRL.123.230501.2019, Levine.PRL.123.170503.2019}. This corresponds to $\Pi_0$ values from 0.78 to 0.94, respectively. As reported in \cite{Omran.Science.365.570.2019}, the fidelity of the generated GHZ state decreases as $N$ increases (specifically, inferred fidelity values are 0.852 for 4 atoms, 0.745 for 8 atoms, 0.643 for 12 atoms, 0.582 for 16 atoms, and 0.542 for 20 atoms). It is expected that the GHZ state fidelity can be increased and generation of even larger GHZ states should be feasible with additional proposed advances \cite{Mitra.PRA.101.030301.2020, Kaubruegger.PRL.123.260505.2019} and technical improvements \cite{Omran.Science.365.570.2019}. However, at this point, it is too early to speculate about how the fidelity will scale with $N$ for very large (100 atoms or more) GHZ states. In any case, the fundamental value of the result of Eq.~\eqref{Delta-phi-imperfect-3} is the demonstration that, as long as the state preparation errors are small, the deviation from the HL is also small.

\section{Effect of laser intensity fluctuations}
\label{sec:intensity-fluctuations}

Random fluctuations in the laser intensity will produce an error in the pulse area:
\begin{equation}
A = \int_{t}^{t+\tau} \Omega_{\mathrm{eff}} (t')  d t' = A_0 + \delta A ,
\end{equation}
where $A_0$ is the nominal (errorless) pulse area and $\delta A$ is the pulse-area error. We assume that $\delta A$ is a random variable that has the wrapped normal distribution with zero mean, whose probability density function is $P(\delta A) = f_{\mathrm{WN}}(\delta A;0,\sigma_{A}^2)$. We denote the errors in the $\pi$ pulse and the $\pi/2$ pulse as $w$ and $v$, respectively. Using Eq.~\eqref{eq:y-rotation}, we obtain the matrices that represent the evolution operators for these pulses:
\begin{align}
& U_{T}^{(k)} (\pi+w) = \begin{bmatrix} 
-S & - e^{i\phi_{T}^{(k)}} C \\[0.2em] e^{-i\phi_{T}^{(k)}} C & -S 
\end{bmatrix} ,  \label{eq:pi-noise} \\
& U_{2T}^{(k)} (\pi/2 + v) = \frac{1}{\sqrt{2}} \begin{bmatrix} 
Q_- & - e^{i\phi_{2T}^{(k)}} Q_+ \\[0.2em] e^{-i\phi_{2T}^{(k)}} Q_+ & Q_- 
\end{bmatrix} ,  \label{eq:pi-over-2-noise}
\end{align}
where we used notation $C \equiv \cos(w/2)$, $S \equiv \sin(w/2)$, and $Q_{\pm} \equiv \cos(v/2) \pm \sin(v/2)$. Using Eqs.~\eqref{eq:pi-noise} and \eqref{eq:pi-over-2-noise}, we obtain the matrix elements of the evolution operator $U_{\text{tot}}^{(k)} = U_{2T}^{(k)} (\pi/2 + v) U_{T}^{(k)} (\pi + w)$:
\begin{subequations}
\label{eq:U-tot-2-pulses-noise} 
\begin{align}
& U_{gg}^{(k)} = - {\textstyle \frac{1}{\sqrt{2}}} \left( Q_- S + e^{i \phi_{2T}^{(k)} - i \phi_{T}^{(k)}} Q_+ C \right) , \\
& U_{ge}^{(k)} = - {\textstyle \frac{1}{\sqrt{2}}} \left( e^{i\phi_{T}^{(k)}} Q_- C - e^{i \phi_{2T}^{(k)}} Q_+ S \right) , \\
& U_{eg}^{(k)} = {\textstyle \frac{1}{\sqrt{2}}} \left( e^{-i\phi_{T}^{(k)}} Q_- C - e^{i\phi_{2T}^{(k)}} Q_+ S \right) , \\
& U_{ee}^{(k)} = - {\textstyle \frac{1}{\sqrt{2}}} \left( Q_- S + e^{i \phi_{T}^{(k)} - i \phi_{2T}^{(k)}} Q_+ C \right) .
\end{align}
\end{subequations}

With the initial GHZ state of Eq.~\eqref{eq:GHZ-state-in} and the evolution operator matrix elements of Eqs.~\eqref{eq:U-tot-2-pulses-noise}, the expectation value of the parity operator is (see Appendix~\ref{sec:app-derivation} for details of the derivation)
\begin{equation}
\braket{\Pi} = \cos^N\! v \cos^{2N} (w/2) \cos\left( \sum_{k=1}^N \phi_k \right) .
\label{eq:Parity-noisy-1} 
\end{equation}
Under the usual assumption of a constant uniform acceleration, we set $\phi_k = \phi = \mathbf{K} \cdot \mathbf{a} T^2$ and $\sum_{k=1}^N \phi_k = N \phi$ in Eq.~\eqref{eq:Parity-noisy-1}. It then follows from Eq.~\eqref{eq:Parity-noisy-1} that the expectation value of the parity operator averaged over the pulse-area error variables, $\overline{\braket{\Pi}}$, has the general form of Eq.~\eqref{eq:Parity-imperfect-avg}, i.e., $\overline{\braket{\Pi}} = \Pi_0 \cos( N \phi )$, where the amplitude of parity oscillations, $\Pi_0$, is given by
\begin{equation}
\Pi_0 = \overline{\cos^N\! v} \, \overline{\cos^{2N} (w/2)}.
\label{eq:Parity-noisy-1a} 
\end{equation}
The averages in Eq.~\eqref{eq:Parity-noisy-1a} are over the wrapped normal distributions with the probability density functions $P(v) = f_{\mathrm{WN}}(v;0,\sigma_v^2)$ and $P(w) = f_{\mathrm{WN}}(w;0,\sigma_w^2)$, which results in
\begin{subequations}
\label{eq:cos-avgs}
\begin{gather}
\overline{\cos^N\! v} = \frac{1}{2^N} \sum_{m = 0}^N \binom{N}{m} e^{-(N-2m)^2 \sigma_v^2 /2} , \\
\overline{\cos^{2N} \frac{w}{2}} = \frac{1}{2^{2N}} \sum_{m = 0}^{2N} \binom{2N}{m} e^{-(N-m)^2 \sigma_w^2 /2} .
\end{gather}
\end{subequations}

The sums in Eqs.~\eqref{eq:cos-avgs} are difficult to compute numerically for large $N$. However, we can approximate them analytically by using an asymptotic formula for the binomial coefficients~\cite{Spencer.book}:
\begin{equation}
\binom{n}{m} \sim \frac{2^n}{\sqrt{n\pi/2}} e^{-(n-2m)^2 /2n} ,
\label{eq:binom-asymptotic}
\end{equation}
which is valid when $n$ is large and $m$ is linear in $n$. By substituting the asymptotic formula \eqref{eq:binom-asymptotic} into Eqs.~\eqref{eq:cos-avgs} and approximating the sums by respective integrals, we obtain an approximate expression for $\Pi_0$, which is valid for $N \gg 1$:
\begin{align}
\Pi_0 \approx \left[ (1 + N \sigma_v^2) (1 + N \sigma_w^2 /2) \right]^{-1/2} .
\label{eq:Parity-noisy-1b}
\end{align}

For the important case of small pulse-area errors such that $N \sigma_v^2 \ll 1$ and $N \sigma_w^2 \ll 1$, we expand the exponentials in Eqs.~\eqref{eq:cos-avgs} to derive:
\begin{equation}
\overline{\cos^N\! v} \approx 1 - \frac{N \sigma_v^2}{2} , \quad
\overline{\cos^{2N} \frac{w}{2}} \approx 1 - \frac{N \sigma_w^2}{4} ,
\end{equation}
which yileds another approximation for $\Pi_0$:
\begin{equation}
\Pi_0 \approx 1 - N \left( \frac{\sigma_v^2}{2} + \frac{\sigma_w^2}{4} \right)   .
\label{eq:Parity-noisy-2} 
\end{equation}

Analogous to the case of a one-dimensional random walk, the variance $\sigma_{A}^2$ is proportional to the pulse duration $\tau$ and, correspondingly, to the nominal pulse area $A_0$. Specifically, $\sigma_v^2 = \xi^2 \pi/2$ and $\sigma_w^2 = \xi^2 \pi$, where the parameter $\xi$ is on the order of $10^{-3}$ or better in state-of-the-art experiments. Therefore, Eq.~\eqref{eq:Parity-noisy-1b} is transformed into
\begin{align}
\Pi_0 \approx (1 + N \xi^2 \pi/2)^{-1} ,
\label{Parity-noisy-1b}
\end{align}
which is valid for $N \gg 1$. Similarly, Eq.~\eqref{eq:Parity-noisy-2} is transformed into
\begin{equation}
\Pi_0 \approx 1 - N \xi^2 \pi/2  ,
\label{eq:Parity-noisy-3} 
\end{equation}
which is valid for $N \xi^2 \pi/2 \ll 1$ (this condition is satisfied for $N \lesssim 10^5$ when $\xi = 10^{-3}$).

\begin{figure}[t]
	\centering
	\includegraphics[width=1.0\linewidth]{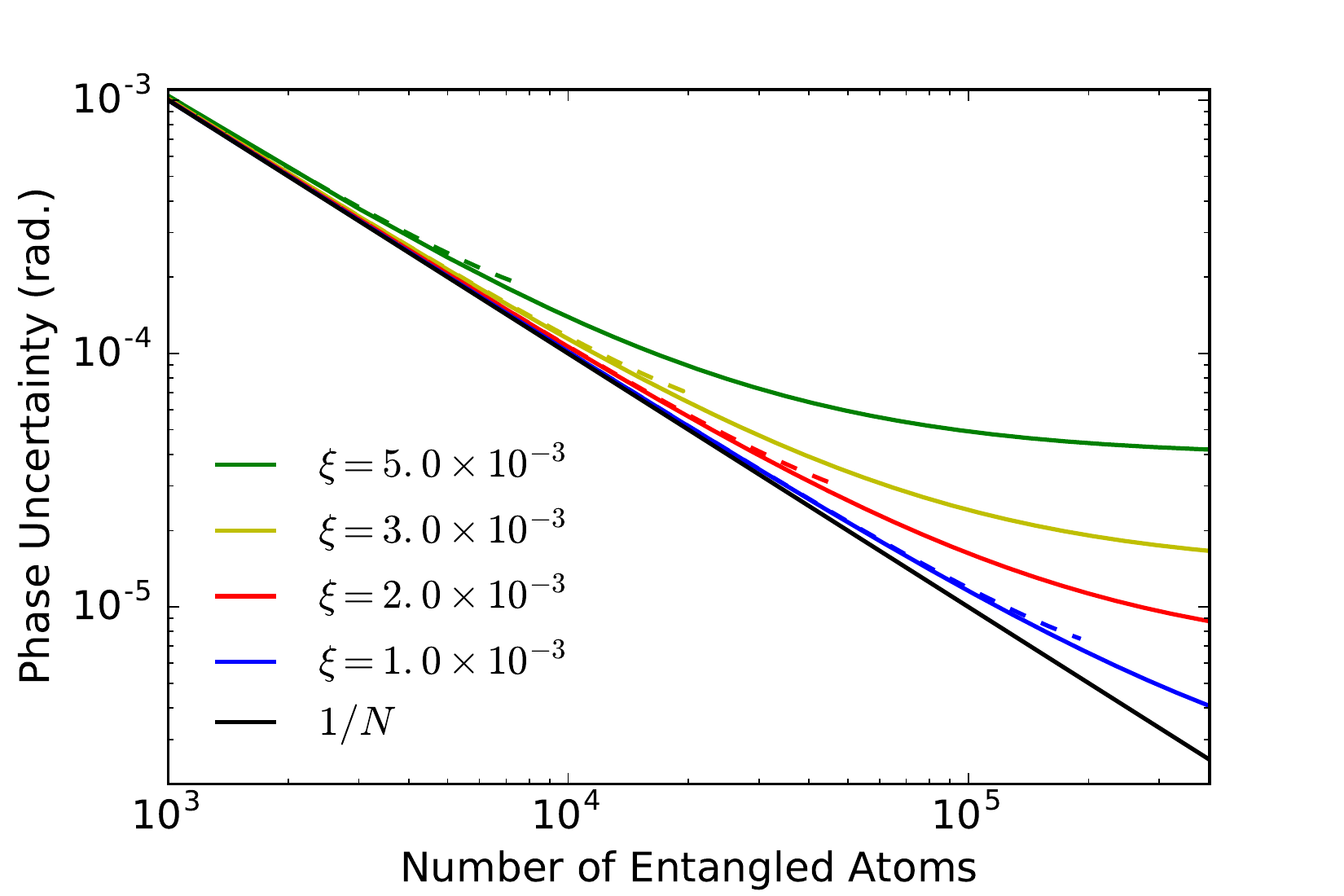}
	\caption{The phase uncertainty $\Delta\phi$ as a function of the number of entangled atoms in the GHZ state, $N$, for various values of the pulse-area error parameter $\xi$. Solid lines show $\Delta\phi$ of Eq.~\eqref{eq:Delta-phi-noise}, which is valid for $N \gg 1$. Dashed lines show $\Delta\phi$ of Eq.~\eqref{eq:Delta-phi-noise-approx}, which is valid for small pulse-area errors; these curves are only shown for values of $N$ such that $N \xi^2 \pi/2 \leq 0.3$. The HL $\Delta\phi = 1/N$ is also shown for comparison.}
	\label{fig:Figure_Phase_Uncertainty_Pulse_Area_Noise}
\end{figure}

The phase uncertainty has the general form of Eq.~\eqref{Delta-phi-imperfect-1} and, for the measurement at a dark fringe, of Eq.~\eqref{Delta-phi-imperfect-2}. The explicit form of the phase uncertainty is, for $N \gg 1$,
\begin{equation}
\label{eq:Delta-phi-noise}
\Delta\phi \approx \frac{1 + N \xi^2 \pi/2}{N} ,
\end{equation}
and, for small pulse-area errors ($N \xi^2 \pi/2 \ll 1$),
\begin{equation}
\label{eq:Delta-phi-noise-approx}
\Delta\phi \approx \frac{1}{(1 - N \xi^2 \pi/2) N} .
\end{equation}
Figure~\ref{fig:Figure_Phase_Uncertainty_Pulse_Area_Noise} shows the phase uncertainty as a function of the number of entangled atoms, $N$, for various values of $\xi$, with solid lines for $\Delta\phi$ of Eq.~\eqref{eq:Delta-phi-noise} and dashed lines for $\Delta\phi$ of Eq.~\eqref{eq:Delta-phi-noise-approx}. The term responsible for the deviation from the HL, $N \xi^2 \pi/2$, increases linearly with $N$. Fortunately, due to the parameter $\xi$ being so small in state-of-the-art experimental conditions, the deviation from the HL is insignificant for at least $N \lesssim 10^5$.

\section{Effect of laser phase noise}
\label{sec:phase-noise}

As shown in Sec.~\ref{sec:background}, the phases of the two Raman fields affect the atomic dynamics only through their difference $\phi_{12} = \phi_1 - \phi_2$. If the two Raman fields are produced by splitting the light from a single laser, the laser phase noise will mostly cancel out in $\phi_{12}$, and the only remaining contribution will be from spurious noise due to technical imperfections, e.g., vibrations and noise in active optical elements such as modulators. While this phase noise could be very small (especially when the two optical paths are well balanced), its effect on the AI performance warrants consideration.

Our analysis is based on using the form~\eqref{eq:U-1} for the evolution operator $U_t (A)$ and assuming $\phi_{12} = -\pi/2 + \vartheta$, where $\vartheta$ is a random phase variable that represents the phase noise. Correspondingly, the evolution operators for the two AI pulses have the following matrix forms:
\begin{align}
& U_{T}^{(k)} (\pi) = \begin{bmatrix} 0 & -e^{i(\phi_{T}^{(k)} + \vartheta_{\pi})} \\ 
e^{-i(\phi_{T}^{(k)} + \vartheta_{\pi})} & 0 \end{bmatrix} ,  \label{eq:pi-phase-noise} \\
& U_{2T}^{(k)} (\pi/2) = \frac{1}{\sqrt{2}} \begin{bmatrix} 1 & -e^{i(\phi_{2T}^{(k)} + \vartheta_{\pi/2})} \\ 
e^{-i(\phi_{2T}^{(k)}  + \vartheta_{\pi/2})} & 1 \end{bmatrix} ,  \label{eq:pi-over-2-phase-noise}
\end{align}
where $\vartheta_{\pi}$ and $\vartheta_{\pi/2}$ denote the random phases in the $\pi$ and $\pi/2$ pulses, respectively. Consequently, the 
evolution operator $U_{\text{tot}}^{(k)} = U_{2T}^{(k)} (\pi/2) U_{T}^{(k)} (\pi)$ has the form of Eq.~\eqref{eq:U-tot-2-pulses} with replacements: $\phi_{T}^{(k)} \rightarrow \phi_{T}^{(k)} + \vartheta_{\pi}$ and $\phi_{2T}^{(k)} \rightarrow \phi_{2T}^{(k)} + \vartheta_{\pi/2}$. Therefore, with the initial GHZ state of Eq.~\eqref{eq:GHZ-state-in}, the expectation value of the parity operator has the form of Eq.~\eqref{parity-EV-GHZ-k} with the replacement $\phi_k \rightarrow \phi_k - \tilde{\vartheta}$, i.e.,
\begin{equation}
\braket{\Pi} = \cos\left[ \sum_{k=1}^N (\phi_k - \tilde{\vartheta}) \right]
= \cos\left( \sum_{k=1}^N \phi_k - N \tilde{\vartheta} \right) ,
\label{eq:Parity-phase-noise-1} 
\end{equation}
where $\tilde{\vartheta} \equiv 2 \vartheta_{\pi} - \vartheta_{\pi/2}$. Assuming, as usual, a constant uniform acceleration, resulting in $\phi_k = \phi = \mathbf{K} \cdot \mathbf{a} T^2$, we obtain:
\begin{equation}
\braket{\Pi} =
\cos\left[ N (\phi - \tilde{\vartheta}) \right] .
\label{eq:Parity-phase-noise-2} 
\end{equation}
Under a general assumption that $\vartheta_{\pi}$ and $\vartheta_{\pi/2}$ are symmetrically distributed around zero, the expectation value of the parity operator in Eq.~\eqref{eq:Parity-phase-noise-2} averaged over the random phase variables has the general form $\overline{\braket{\Pi}} = \Pi_0 \cos( N \phi )$, where \begin{equation}
\Pi_0 = \overline{\cos(N \tilde{\vartheta})} .
\label{eq:Parity-phase-noise-2a} 
\end{equation}
Next, we assume that each of the random phase variables $\vartheta_{\pi}$ and $\vartheta_{\pi/2}$ has the same distribution, which is the wrapped normal distribution with zero mean and variance $\sigma_{\vartheta}^2$, whose probability density function is $P(\vartheta_{\mu}) = f_{\mathrm{WN}}(\vartheta_{\mu};0,\sigma_{\vartheta}^2)$, where $\mu = \{ \pi,\pi/2 \}$. Then $\tilde{\vartheta}$ is also a random phase variable that has the wrapped normal distribution with $P(\tilde{\vartheta}) = f_{\mathrm{WN}}(\tilde{\vartheta};0, r_{\text{corr}} \sigma_{\vartheta}^2)$, where the value of the numerical factor $r_{\text{corr}}$ depends on the degree of correlation between $\vartheta_{\pi}$ and $\vartheta_{\pi/2}$. This correlation is determined by the ratio between the characteristic time scale of the spurious noise and the time separation $T$ between the $\pi$ and $\pi/2$ pulses. If $\vartheta_{\pi}$ and $\vartheta_{\pi/2}$ are completely correlated then $r_{\text{corr}} = 1$, if $\vartheta_{\pi}$ and $\vartheta_{\pi/2}$ are completely independent then $r_{\text{corr}} = 5$, and in general $1 \leq r_{\text{corr}} \leq 5$. With this assumption, we obtain:
\begin{equation}
\Pi_0 = \exp(- N^2 \sigma_{\vartheta}^2 r_{\text{corr}}/2) \approx 1 - N^2 \sigma_{\vartheta}^2 r_{\text{corr}}/2 ,
\label{eq:Parity-phase-noise-3} 
\end{equation}
where the approximate expression is valid for small phase errors such that $N^2 \sigma_{\vartheta}^2 \ll 1$. 

\begin{figure}[htbp]
	\centering
	\includegraphics[width=1.0\linewidth]{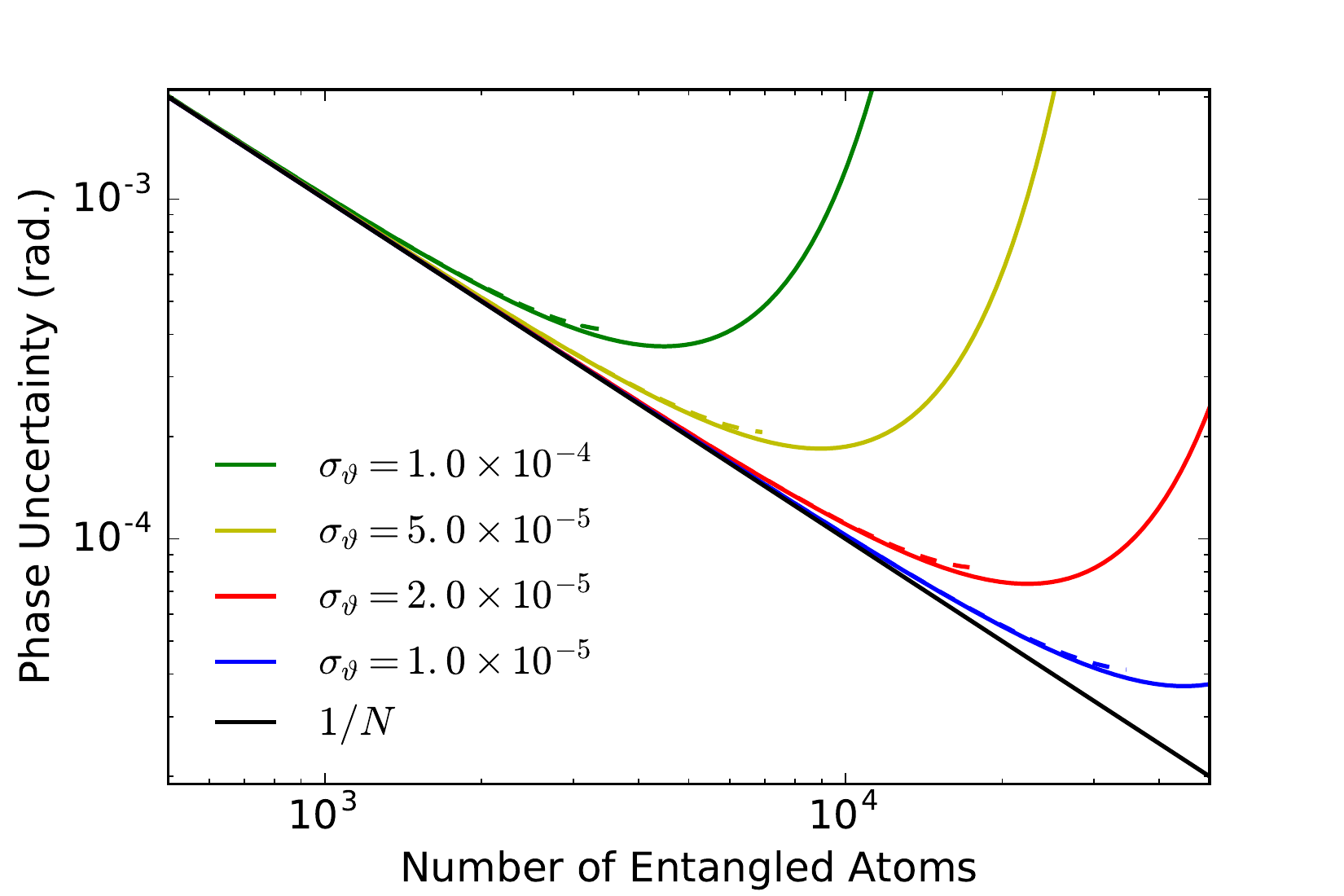}
	\caption{The phase uncertainty $\Delta\phi$ as a function of the number of entangled atoms in the GHZ state, $N$, for various values of the standard deviation of the phase error, $\sigma_{\vartheta}$, and $r_{\text{corr}} = 5$. Solid lines show $\Delta\phi$ of Eq.~\eqref{eq:Delta-phi-phase-noise}, which is the exact result. Dashed lines show $\Delta\phi$ of Eq.~\eqref{eq:Delta-phi-phase-noise-approx}, which is an approximation valid for small phase errors; these curves are only shown for values of $N$ such that $N^2 \sigma_{\vartheta}^2 r_{\text{corr}}/2 \leq 0.3$. The HL $\Delta\phi = 1/N$ is also shown for comparison.}
	\label{fig:Figure_Phase_Uncertainty_Phase_Noise}
\end{figure}

Once again, the phase uncertainty has the general form of Eq.~\eqref{Delta-phi-imperfect-1} and, for the measurement at a dark fringe, of Eq.~\eqref{Delta-phi-imperfect-2}. The explicit form of the phase uncertainty is
\begin{align}
\Delta\phi = \frac{1}{\exp(- N^2 \sigma_{\vartheta}^2 r_{\text{corr}}/2)  N}  \label{eq:Delta-phi-phase-noise} \\
\approx \frac{1}{(1 - N^2 \sigma_{\vartheta}^2 r_{\text{corr}}/2) N} , \label{eq:Delta-phi-phase-noise-approx}
\end{align}
where the approximation in Eq.~\eqref{eq:Delta-phi-phase-noise-approx} is valid for small phase errors ($N^2 \sigma_{\vartheta}^2 \ll 1$).
Figure~\ref{fig:Figure_Phase_Uncertainty_Phase_Noise} shows the phase uncertainty as a function of the number of entangled atoms, $N$, for various values of $\sigma_{\vartheta}$ and $r_{\text{corr}} = 5$, with solid lines for $\Delta\phi$ of Eq.~\eqref{eq:Delta-phi-phase-noise} and dashed lines for $\Delta\phi$ of Eq.~\eqref{eq:Delta-phi-phase-noise-approx}.
The term responsible for the deviation from the HL, $N^2 \sigma_{\vartheta}^2 r_{\text{corr}}/2$, increases quadratically with $N$. While we do not know the exact value of $\sigma_{\vartheta}$, it should be very small based on considerations described above, which ensures that the deviation from the HL is insignificant even for large values of $N$.

\section{Effect of initial momentum distribution}
\label{sec:Doppler-shift}

In this section we investigate the effect that the momentum uncertainty of the atoms has on the phase uncertainty of the AI. This momentum uncertainty arises because each atom is initially trapped in individual optical tweezers. In the anticipated AI protocol, the tweezers are extinguished, and short pulses of the Raman and Rydberg lasers generate the GHZ spin sate, while also imparting a state-dependent momentum kick to the atoms. The GHZ state preparation time can be made as short as $\sim 1~\mu\mathrm{s}$ using optimally shaped pulses \cite{Omran.Science.365.570.2019}. Hence we can neglect atomic motion during this step. After the GHZ state has been prepared, the first free evolution step of the AI operation begins, and we assume that the atoms are in the same motional state in which they existed in the traps, plus the state-dependent momentum kick from the entangling procedure.

\subsection{Description of the trapped atoms}

The trapped atoms are confined in three dimensions and have three components of vibrational motion, but the only motion relevant to the interferometer is that along the direction of the momentum kick $\hbar \mathbf{K}$ from the two-photon stimulated Raman transition. This is clear because the atomic momentum $\mathbf{p}$ always appears in the dot product $\mathbf{p} \cdot \mathbf{K}$ in the effective Hamiltonian for the Raman transition, as shown in Eq.~\eqref{eq:delta12}. Therefore, we consider motion of the atoms in only one dimension, along the coordinate parallel to $\mathbf{K}$. We denote the position and momentum of the $k$th atom along this coordinate as $x_k$ and $p_k$, respectively.

We assume that each atom is initially in its own harmonic trap with energy eigenstates $\ket{n_k}$, $E_{n_k} = (n_k+\frac{1}{2}) \hbar\omega_{\mathrm{trap}}$, $n_k = \{ 0,1,2,\ldots \}$, where $\omega_{\mathrm{trap}} = 2\pi \nu_{\mathrm{trap}}$ is the trap frequency which is assumed to be identical for all traps. The position-space and momentum-space representations of the state $\ket{n_k}$ are, respectively:
\begin{subequations}
\begin{align}
& \psi_{n_k}(x_k) = \left( 2^{n_k} n_k! \sqrt{\pi \sigma_x^2} \right)^{-1/2} e^{-x_k^2/ 2\sigma_x^2} H_{n_k}( x_k/\sigma_x ),  \label{eq:psi_x} \\
& \tilde{\psi}_{n_k}(p_k) = i^{n_k} \left( 2^{n_k} n_k! \sqrt{\pi \sigma_p^2} \right)^{-1/2} e^{-p_k^2/ 2 \sigma_p^2} H_{n_k}(p_k/\sigma_p),   \label{eq:psi_p}
\end{align}
\end{subequations}
where $H_{n_k}$ is the Hermite polynomial of degree $n_k$. The parameter $\sigma_x = \sqrt{\hbar/ m\omega_{\mathrm{trap}}}$ is the length scale of the trap, and $\sigma_p = \hbar/\sigma_x = \sqrt{\hbar m\omega_{\mathrm{trap}}}$, where $m$ is the atom's mass. In this paper, we consider $^{133}$Cs atoms for which $m \approx 132.90545$~u. The position and momentum uncertainties of the ground vibrational state (for one atom) are $(\Delta x)_0 = \sigma_x/\sqrt{2}$ and $(\Delta p)_0 = \sigma_p/\sqrt{2}$.

The vibrational motion of the trapped atoms is decoupled from their spin state. As a result, the total density matrix of the atoms has the form:
\begin{equation}
\rho =  \rho_{\mathrm{spin}} \otimes \rho_{\mathrm{vib}} .
\end{equation}
We assume that each atom is in a thermal vibrational state, with temperature $\mathcal{T}$, identical for all atoms. The thermal vibrational state of $N$ atoms has the form:
\begin{equation}
\rho_{\mathrm{vib}} = \bigotimes_{k=1}^N \rho_{\mathrm{vib}}^{(k)} = 
\bigotimes_{k=1}^N \sum_{n_k = 0}^{\infty} \frac{z^{n_k}}{1+\braket{n}} \ket{n_k}\bra{n_k} , 
\label{eq:vibration}
\end{equation}
where $\braket{n} = [\exp(\hbar \omega_{\mathrm{trap}}/k_{\mathrm{B}} \mathcal{T}) - 1]^{-1}$ is the average vibrational excitation number, which is identical for all atoms, and $z \equiv \braket{n}/(1+\braket{n}) = \exp(-\hbar \omega_{\mathrm{trap}}/k_{\mathrm{B}} \mathcal{T})$.

As the spins of the atoms can be entangled, we express the spin density matrix in terms of the general $N$-atom spinor $X_{\vec{\alpha}}$:
\begin{equation}
\rho_{\mathrm{spin}} = \sum_{\vec{\alpha},\vec{\alpha}'} X_{\vec{\alpha}} X_{\vec{\alpha}'} \ket{\vec{\alpha}} \bra{\vec{\alpha}'} ,
\end{equation}
where $\vec{\alpha}$ stands for the set of indices $\{ \alpha_1, \alpha_2, \ldots, \alpha_N \}$ and $\alpha_k = \{g,e\}$.

Immediately after the traps are switched off, the atoms undergo an entangling process that prepares them in the GHZ spin state and, due to the use of counter-propagating Raman beams, provides a momentum kick of $\hbar K$, where $K = |\mathbf{K}| \approx 4\pi/\lambda_{\text{D}_2} \approx 1.4743 \times 10^7$~$\text{m}^{-1}$, to the excited spin state only. In the GHZ state, the components of $X_{\vec{\alpha}}$ acquire the following values: $X_{\vec{\alpha}} = 1/\sqrt{2}$ if $\alpha_k = g$ for all $k$, $X_{\vec{\alpha}} = 1/\sqrt{2}$ if $\alpha_k = e$ for all $k$, and $X_{\vec{\alpha}} = 0$ otherwise. The spin-dependent momentum kick for the $k$th atom is denoted as $\hbar K_{\alpha_k}$, where $K_e = K$ and $K_g = 0$.

To represent the effect of this entangling procedure, we modify the density matrix $\rho_{\mathrm{vib}}$ in Eq.~\eqref{eq:vibration} by expanding the energy eigenstates $\ket{n_k}$ in the basis of momentum eigenstates and shifting their momenta by $\hbar K_{\alpha_k}$. Hence we can express the total density matrix at $t=0$ as
\begin{align}
\rho(0) =\, & \sum_{\vec{\alpha},\vec{\alpha}'} X_{\vec{\alpha}} X_{\vec{\alpha}'} \bigotimes_{k=1}^N 
\int_{-\infty}^{\infty} d p_k  \int_{-\infty}^{\infty} d p'_k \, P(p_k,p'_k) \nonumber \\
& \times \ket{\alpha_k,p_k + \hbar K_{\alpha_k}} \bra{\alpha'_k,p'_k + \hbar K_{\alpha'_k}} , 
\label{eq:rho_intial_2}
\end{align}
where $P(p_k,p'_k) \equiv \bra{p_k} \rho_{\mathrm{vib}}^{(k)} \ket{p'_k}$ is given by
\begin{equation}
P(p_k,p'_k) = \sum_{n_k = 0}^{\infty} \frac{z^{n_k}}{1+\braket{n}} 
\tilde{\psi}_{n_k}(p_k) \tilde{\psi}^{\ast}_{n_k}(p'_k) .
\label{eq:S-definition}
\end{equation}
The sum in Eq.~\eqref{eq:S-definition} can be evaluated analytically. Using Eq.~(\ref{eq:psi_p}), we rewrite $P(p_k,p'_k)$ as
\begin{equation}
P(p_k,p'_k) = \frac{e^{-(p_k^2+{p'_k}^2)/2\sigma_p^2} }{(1+\braket{n})\sqrt{\pi} \sigma_p} G(p_k,p'_k;z)
\label{eq:S-1}
\end{equation}
where
\begin{align}
G(p_k,p'_k;z) & = \sum_{n_k = 0}^{\infty} \frac{(z/2)^{n_k}}{n_k!} 
H_{n_k}(p_k/\sigma_p) H_{n_k}(p'_k/\sigma_p) \nonumber \\
& = \frac{ e^{[2 p_k p'_k z - (p_k^2+{p'_k}^2) z^2 ]/[\sigma_p^2 (1-z^2)]} }{\sqrt{1-z^2}} 
\label{eq:Hermite-GF}
\end{align}
is the generating function for the Hermite polynomials~\cite{Erdelyi.book}. Substituting Eq.~\eqref{eq:Hermite-GF} into Eq.~\eqref{eq:S-1}, we derive the following expression for $P(p_k,p'_k)$:
\begin{equation}
\label{eq:S}
P(p_k,p'_k) = \frac{e^{[ 4 p_k p'_k z - (p_k^2+{p'_k}^2) (1+z^2) ]/[2 \sigma_p^2 (1-z^2)] }  
}{\sqrt{ (2 \braket{n}+1) \pi} \sigma_p} .
\end{equation}
Note that $2 \braket{n} + 1 = (1+z)/(1-z) = \coth (\hbar \omega_{\mathrm{trap}}/2 k_{\mathrm{B}} \mathcal{T})$ and $(1+z^2)/(1-z^2) = \coth (\hbar \omega_{\mathrm{trap}}/k_{\mathrm{B}} \mathcal{T})$. 

\subsection{Parity expectation value}

Analogously to Eq.~\eqref{eq:Parity-EV-general}, the expectation value of the parity operator is
\begin{equation}
\label{eq:Parity-EV-general-rho}
\braket{\Pi} = \text{Tr}\left[ U_{\text{tot}} \rho(0) U_{\text{tot}}^{\dagger}  (\Pi \otimes \openone_{\mathbf{p}}) \right] ,
\end{equation} 
where $\rho(0)$ is the initial density matrix, given by Eq.~\eqref{eq:rho_intial_2}, and $U_{\text{tot}} = \bigotimes_{k=1}^N U_{\text{tot}}^{(k)}$ is the evolution operator for the AI operation. With the initial GHZ state, the AI performs the $\pi$--$\pi/2$ pulse sequence, and therefore $U_{\text{tot}}^{(k)}$ is given by
\begin{equation}
U_{\text{tot}}^{(k)} = U_{2T}^{(k)} (\tau_{\pi}/2) U_{T}^{(k)} (\tau_{\pi}) ,
\label{eq:U-2-pulses-with-detunerr}
\end{equation}
where the evolution operator for each of the pulses, $U_{t}^{(k)}(\tau)$, is given by Eq.~\eqref{eq:U-general}, where $t = T$, $\tau = \tau_{\pi}$ for the $\pi$ pulse and $t = 2T$, $\tau = \tau_{\pi}/2$ for the $\pi/2$ pulse.

By substituting the initial density matrix $\rho(0)$ of Eq.~\eqref{eq:rho_intial_2} into Eq.~\eqref{eq:Parity-EV-general-rho}, we obtain:
\begin{align}
\braket{\Pi} =\, & \sum_{\vec{\alpha},\vec{\alpha}'} X_{\vec{\alpha}} X_{\vec{\alpha}'} \bigotimes_{k=1}^N 
\int_{-\infty}^{\infty} d p_k  P(p_k) \nonumber \\
& \times \bra{\alpha'_k,p_k + \hbar K_{\alpha'_k}} U_{\text{tot}}^{(k) \dagger} \Pi U_{\text{tot}}^{(k)} \ket{\alpha_k,p_k + \hbar K_{\alpha_k}} , 
\label{eq:Parity-EV-general-rho-1}
\end{align}
where $P(p_k) = P(p_k,p_k) = \bra{p_k} \rho_{\mathrm{vib}}^{(k)} \ket{p_k}$ is the momentum distribution for the thermal vibrational state of one atom. Using Eq.~\eqref{eq:S}, we obtain:
\begin{equation}
\label{eq:S-final}
P(p_k) = \frac{e^{- p_k^2 /2 \sigma_{\text{th}}^2 } }{\sqrt{2 \pi} \sigma_{\text{th}} } ,
\end{equation}
which is the probability density function of the normal distribution for variable $p_k$, with zero mean and variance $\sigma_{\text{th}}^2$, where $\sigma_{\text{th}} \equiv (\Delta p)_{\text{th}} = \sqrt{\braket{p^2}} = \sigma_p \sqrt{\braket{n} + 1/2}$ is the momentum uncertainty of the thermal state for one atom.

In contrast to the analysis in Secs.~\ref{sec:ideal}--\ref{sec:phase-noise}, in this section we do not make the approximation used to obtain Eq.~\eqref{eq:U-1}, i.e., we do not neglect terms on the order of $|\delta_{12} - \delta^{\mathrm{AC}}| / \Omega_\text{eff}$ and $|\delta_{12}| \tau$ in Eq.~\eqref{eq:U-general}. As mentioned above, in practice, laser frequency chirping is used to compensate the evolving Doppler shift due to the acceleration of the atom, and thereby keep $|\delta_{12}|$ small enough for this approximation to be valid~\cite{Kasevich.APB.54.321.1992, Young.chapter.1997}. However, if the momentum spread of the initial state $\rho(0)$ is significant, then it might be impossible to make the aforementioned terms negligible for all momentum components of the atomic wave packet simultaneously. As we mentioned above, in the frame that accelerates with the atom, the Raman frequency experiences a time-dependent Doppler shift: $\omega_{12}(t) = \omega_{12}(0) - K a_k t$, where $a_k = \mathbf{a}_k \cdot \mathbf{K}/K$ is the component of $\mathbf{a}_k$ parallel to $\mathbf{K}$. If one also implements a linear chirp $b t$ of the Raman frequency $\omega_{12}(t)$, the full time dependence is $\omega_{12}(t) = \omega_{12}(0) + (b - K a_k) t$ \cite{Kasevich.APB.54.321.1992}. Correspondingly, an expression for the Raman detuning that explicitly takes into account the time dependence is 
\begin{equation}
\delta_{12}^{(k)}(t) = \omega_{12}(0) - \omega_{eg} - \frac{p_k K}{m} - \frac{\hbar K^2}{2 m} + (b - K a_k) t . 
\end{equation}
It is customary to set the Raman frequency at $t = 0$ to be at the resonance for $p_k = 0$, i.e.,
\begin{equation}
\omega_{12}(0) = \omega_{eg} + \frac{\hbar K^2 }{2m} + \delta^{\mathrm{AC}} ,
\label{eq:resonance-condition}
\end{equation}
which yields
\begin{equation}
{\delta'}_{12}^{(k)}(t) \equiv \delta_{12}^{(k)}(t) - \delta^{\mathrm{AC}} =  - \frac{p_k K}{m} + (b - K a_k) t . 
\label{eq:delta12-at-resonance}
\end{equation}
The relative ac Stark shift $\delta^{\mathrm{AC}}$ can be tuned to zero by adjusting the intensities and polarizations of the Raman fields~\cite{Wineland.PTRSLA.361.1349.2003}, but we include it in Eqs.~\eqref{eq:resonance-condition} and \eqref{eq:delta12-at-resonance} for the sake of generality. In this section, we assume that the value of the frequency chirp rate $b$ is chosen such that the term $(b - K a_k) t$ is much smaller than $\sigma_{\mathrm{th}} K/m$ at the times when the pulses are applied (i.e., for $t = T$ and $t = 2T$), and therefore we set 
\begin{equation}
{\delta'}_{12}^{(k)} = \delta_{12}^{(k)} - \delta^{\mathrm{AC}} = - \frac{p_k K}{m}. 
\label{eq:delta12-at-resonance-simplified}
\end{equation}
We consider the effect of the relatively small corrections to the detuning, $(b - K a_k) T$ and $2 (b - K a_k) T$, in Appendix~\ref{sec:app-corrections}. We also make the choice $\phi_{12} + \delta^{\mathrm{AC}} \tau/2 = -\pi/2$ for each of the pulses.

By using the evolution operator of Eq.~\eqref{eq:U-2-pulses-with-detunerr} in Eq.~\eqref{eq:Parity-EV-general-rho-1}, the expectation value of the parity operator is (see Appendix~\ref{sec:app-derivation} for details of the derivation)
\begin{equation}
\braket{\Pi} = (1-\eta)^N \cos\left( \sum_{k=1}^N \phi_k \right) ,
\label{eq:app:Parity-EV-entangled-mse-3} 
\end{equation}
where 
\begin{equation}
\eta \equiv 1 - \int_{-\infty}^{\infty}\! d p P(p) \frac{\sin^3 2\lambda}{(1+r^2)^{3/2}} 
\left[ \cos \pi r + \frac{r \tan\lambda \sin \pi r}{\sqrt{1+r^2}} \right]  \label{eq:eta-general}
\end{equation}
and
\begin{equation}
r \equiv \frac{\delta'_{12}}{\Omega_{\text{eff}}} = - \frac{p K}{m \Omega_{\text{eff}}} , \quad
\lambda \equiv \frac{\pi}{4} \sqrt{1+r^2} .
\end{equation}
Under the usual assumption of a constant uniform acceleration, i.e., $\phi_k = \phi = (b - \mathbf{K} \cdot \mathbf{a}) T^2$, the parity expectation value of Eq.~\eqref{eq:app:Parity-EV-entangled-mse-3} has the general form $\braket{\Pi} = \Pi_0 \cos( N \phi )$, where $\Pi_0 = (1-\eta)^N$. In the regime where $N \eta \ll 1$, we can use the approximation $\Pi_0 \approx 1- N \eta$. 

The value of $\eta$ can be easily computed via numerical integration in Eq.~\eqref{eq:eta-general}. However, it is instructive to obtain an approximate expression for $\eta$, which is valid when the uncertainty of the Doppler shift term, $(\Delta \delta'_{12})_{\text{th}} = K \sigma_{\text{th}}/m$, is small compared to $\Omega_{\text{eff}}$. We expand the integrand in Eq.~\eqref{eq:eta-general} in the powers of $r^2$ and neglect all terms on the order of $r^4$ or smaller, to obtain:
\begin{equation}
\label{eq:eta-approx-1}
\eta \approx \kappa \braket{r^2} 
= \kappa \left(\frac{K}{m \Omega_{\text{eff}}} \right)^2 \! \braket{p^2} 
= \kappa \left(\frac{K \sigma_{\text{th}}}{m \Omega_{\text{eff}}} \right)^2 ,
\end{equation}
where $\kappa = {\textstyle \frac{1}{2}}(\pi^2 + 3 - 2\pi) \approx 3.2932$ is a numerical factor. Using the explicit form $\sigma_{\text{th}}^2 = \hbar m\omega_{\mathrm{trap}}(\braket{n} + 1/2)$, we rewrite Eq.~\eqref{eq:eta-approx-1} as
\begin{equation}
\label{eq:eta-approx-2}
\eta \approx \kappa \frac{K^2 \braket{E_{\text{vib}}} }{m \Omega_{\mathrm{eff}}^2} ,
\end{equation}
where
\begin{equation}
\label{eq:E-vib}
\braket{E_{\text{vib}}} = \hbar \omega_{\mathrm{trap}} \left(\braket{n} + {\textstyle \frac{1}{2}}\right)
\end{equation}
is the average vibrational energy of the atom in the trap. In the limit of high temperature, $\hbar \omega_{\mathrm{trap}} \ll k_{\mathrm{B}} \mathcal{T}$, we have $\braket{E_{\text{vib}}} \approx k_{\mathrm{B}} \mathcal{T}$, and $\eta$ is independent of the trap frequency. In the limit of low temperature, $\hbar \omega_{\mathrm{trap}} \gg k_{\mathrm{B}} \mathcal{T}$, we have $\braket{E_{\text{vib}}} \approx \hbar \omega_{\mathrm{trap}}/2$ (the ground state energy), and $\eta$ scales linearly with the trap frequency. 

The phase uncertainty has the general form of Eq.~\eqref{Delta-phi-imperfect-1} and, for the measurement at a dark fringe, of Eq.~\eqref{Delta-phi-imperfect-2}. The explicit form of the phase uncertainty is
\begin{subequations}
\begin{align}
\Delta\phi & = \frac{1}{(1 - \eta)^N N}  \label{eq:Delta-phi-mse-num} \\
& \approx \frac{1}{(1 - \kappa K^2 \braket{E_{\text{vib}}} /m \Omega_{\mathrm{eff}}^2)^N N} . \label{eq:Delta-phi-mse-approx}
\end{align}
\end{subequations}

From Eq.~\eqref{eq:eta-approx-2}, the strategy to minimize the error associated with the momentum uncertainty is very straightforward: minimize the vibrational energy by cooling the atoms as close as possible to the ground state and lowering the trap frequency, and increase the Rabi frequency for the Raman transition by using a high-intensity laser with tight focusing. Ideally, we would prefer the regime in which atoms are cooled to sub-microkelvin temperatures, the trap frequency is lowered to $\nu_{\mathrm{trap}} < 10$~kHz, and the Rabi frequency is increased to $\Omega_{\mathrm{eff}} > 2\pi \times 500$~kHz. A recent experiment~\cite{Levine.PRL.123.170503.2019} reported a Rabi frequency of $\Omega_{\mathrm{eff}} \approx 2\pi \times 250$~kHz for a two-photon Raman transition driven by a laser field tuned near the D$_1$ ($5^2\text{S}_{1/2} \longrightarrow 5^2\text{P}_{1/2}$) transition of $^{87}$Rb. Since the dipole moment for the D$_2$ ($6^2\text{S}_{1/2} \longrightarrow 6^2\text{P}_{3/2}$) transition of $^{133}$Cs is about 1.5 times larger than that for the D$_1$ transition of $^{87}$Rb, one can expect that a laser system with the same intensity and focusing as the one used in Ref.~\cite{Levine.PRL.123.170503.2019} would produce a Rabi frequency of $\Omega_{\mathrm{eff}} \approx 2\pi \times 560$~kHz for a two-photon Raman transition driven by a laser field tuned near the D$_2$ transition of $^{133}$Cs.

\begin{figure}[htbp]
	\centering
	\includegraphics[width=0.98\linewidth]{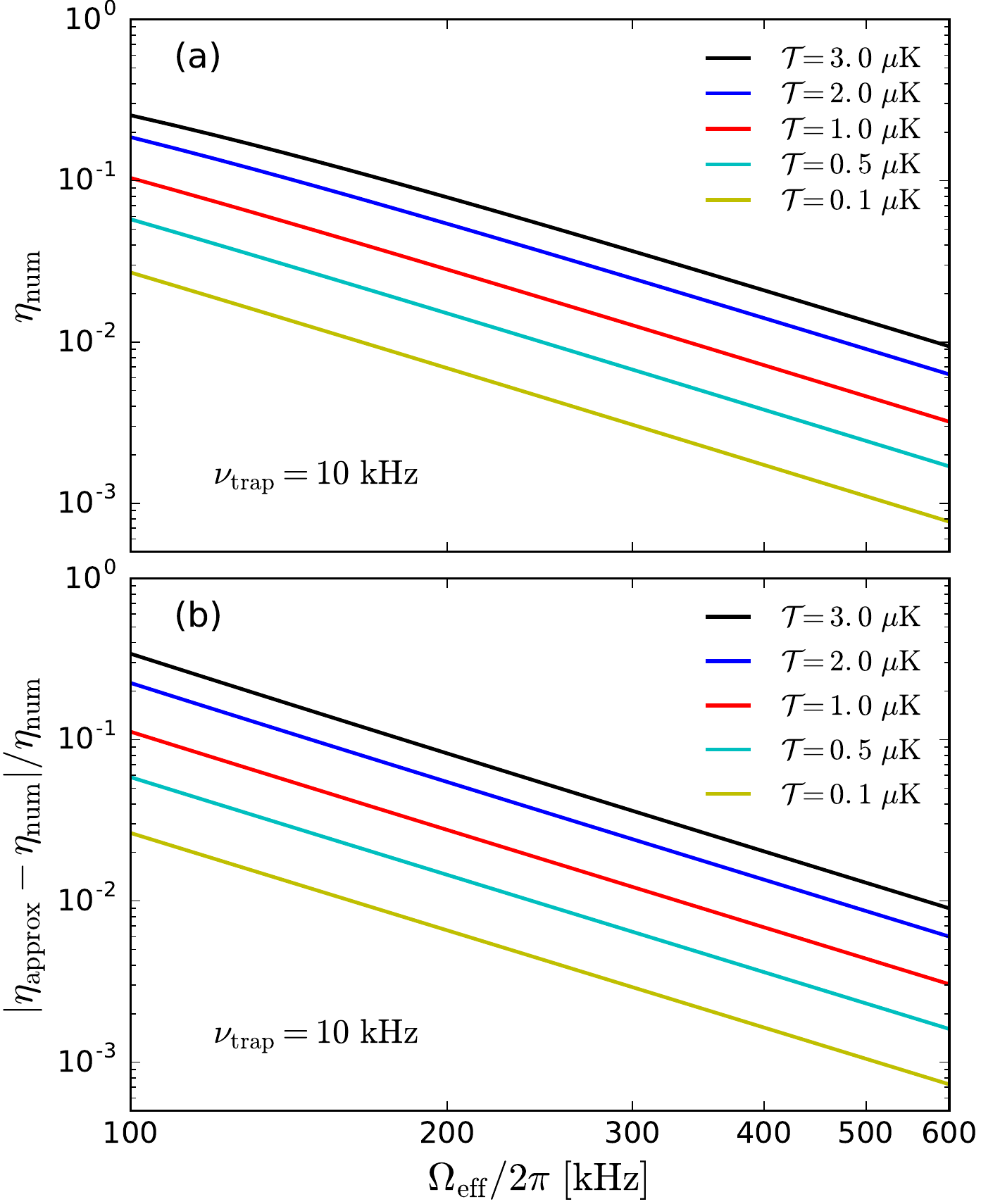}
	\caption{(a) $\eta_{\text{num}}$ and (b) $|\eta_{\mathrm{approx}} - \eta_{\mathrm{num}}|/\eta_{\mathrm{num}}$, as functions of the Rabi frequency $\Omega_{\mathrm{eff}} / 2\pi$, for various values of the atom temperature $\mathcal{T}$ and a fixed value of the trap frequency ($\nu_{\mathrm{trap}} = 10$~kHz).}
	\label{fig:eta_vs_RabiFreq}
\end{figure}

\begin{figure}[htbp]
	\centering
	\includegraphics[width=0.98\linewidth]{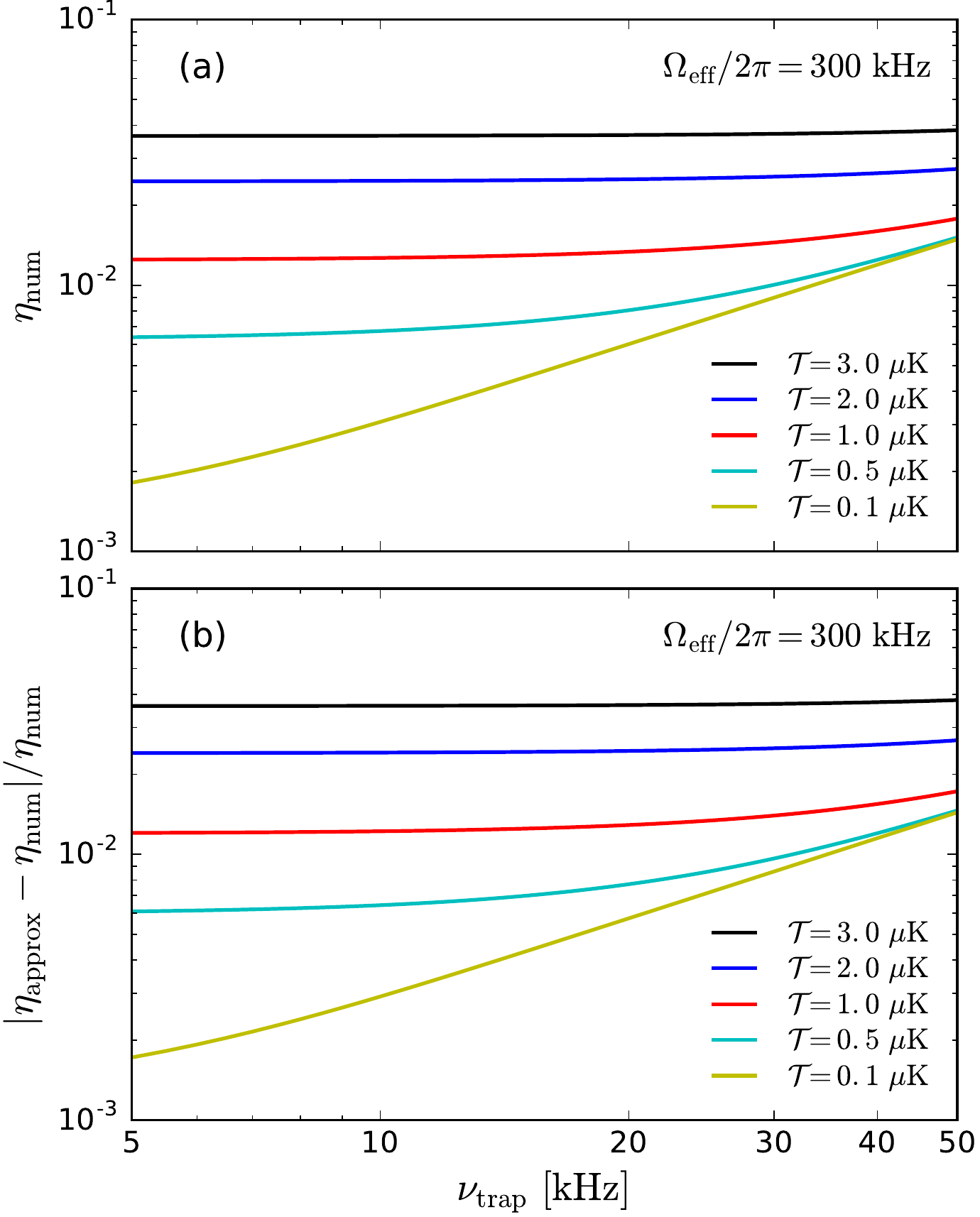}
	\caption{(a) $\eta_{\text{num}}$ and (b) $|\eta_{\mathrm{approx}} - \eta_{\mathrm{num}}|/\eta_{\mathrm{num}}$, as functions of the trap frequency $\nu_{\mathrm{trap}}$, for various values of the atom temperature $\mathcal{T}$ and a fixed value of the Rabi frequency ($\Omega_{\mathrm{eff}}/2\pi = 300$~kHz).}
	\label{fig:eta_vs_nu}
\end{figure}

\subsection{Numerical results}

To differentiate between exact (numerical) and approximate (analytical) results, we denote the values of $\eta$ obtained via numerical integration in Eq.~\eqref{eq:eta-general} as $\eta_{\text{num}}$ and the values calculated using the approximate analytical formula in Eq.~\eqref{eq:eta-approx-2} as $\eta_{\text{approx}}$. The relative error of the approximate value of $\eta$ is $|\eta_{\mathrm{approx}} - \eta_{\mathrm{num}}|/\eta_{\mathrm{num}}$. Similarly, we denote the values of the phase uncertainty obtained by substituting $\eta_{\text{num}}$ into Eq.~\eqref{eq:Delta-phi-mse-num} as $(\Delta\phi)_{\mathrm{num}}$ and the values calculated using the approximate analytical formula in Eq.~\eqref{eq:Delta-phi-mse-approx} as $(\Delta\phi)_{\mathrm{approx}}$. The relative error of the approximate value of the phase uncertainty is $|(\Delta\phi)_{\mathrm{approx}} - (\Delta\phi)_{\mathrm{num}}|/(\Delta\phi)_{\mathrm{num}}$.

\begin{figure}[htbp]
	\centering
	\includegraphics[width=0.98\linewidth]{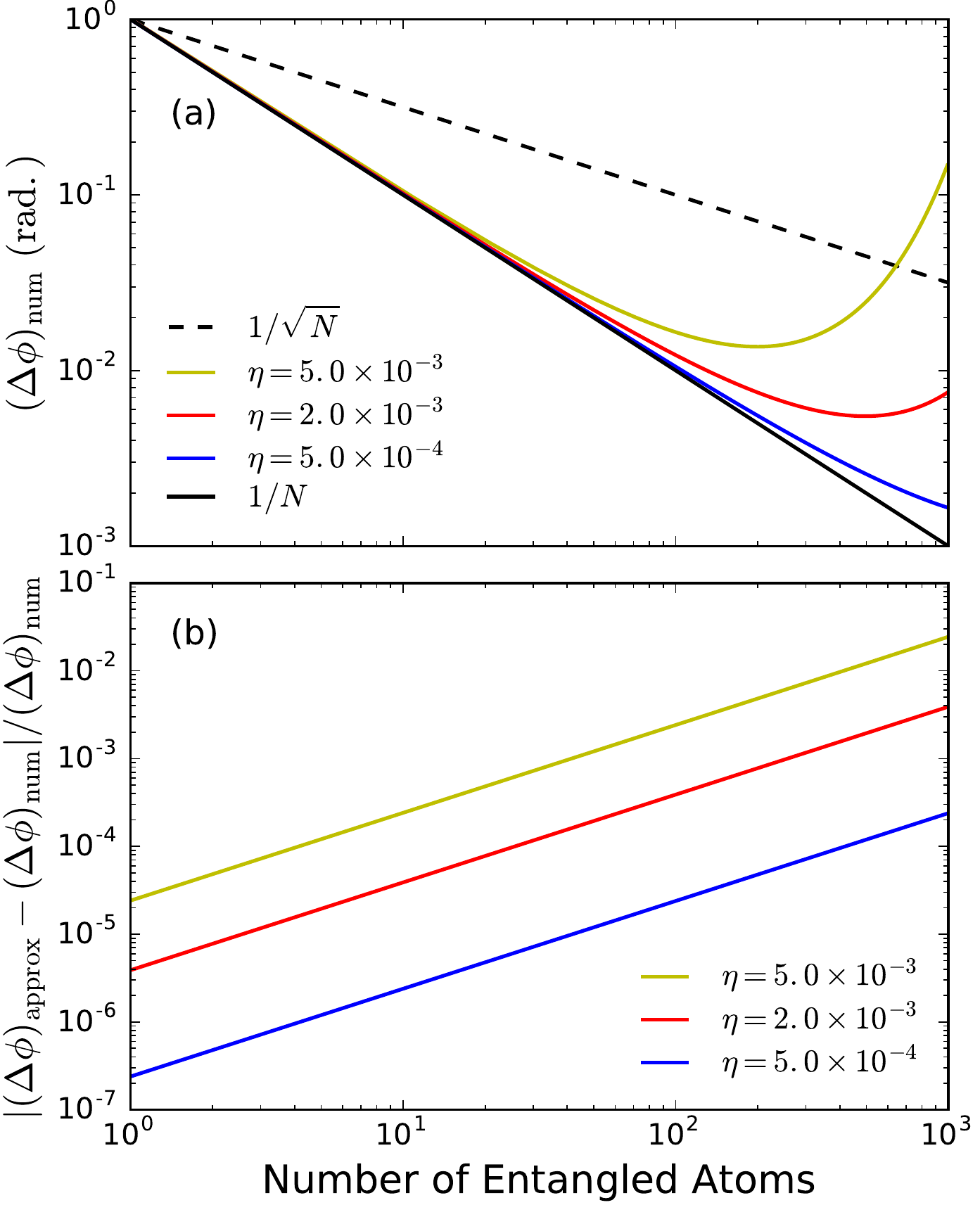}
	\caption{(a) The phase uncertainty $(\Delta\phi)_{\mathrm{num}}$ and (b) the relative error $|(\Delta\phi)_{\mathrm{approx}} - (\Delta\phi)_{\mathrm{num}}|/(\Delta\phi)_{\mathrm{num}}$, as functions of the number of entangled atoms in the GHZ state, $N$, for three parameter combinations (described in Table~\ref{table:curve_parameters}) with various values of $\eta$. In subplot (a), the HL $\Delta\phi = 1/N$ and the SQL $\Delta\phi = 1/N^{1/2}$ are also shown for comparison.}
	\label{fig:Dphi_vs_N}
\end{figure}

Figure~\ref{fig:eta_vs_RabiFreq} shows $\eta_{\text{num}}$ and $|\eta_{\mathrm{approx}} - \eta_{\mathrm{num}}|/\eta_{\mathrm{num}}$ as functions of the Rabi frequency $\Omega_{\mathrm{eff}} / 2\pi$ for various values of the atom temperature $\mathcal{T}$ and a fixed value of the trap frequency ($\nu_{\mathrm{trap}} = 10$~kHz). As expected from Eq.~\eqref{eq:eta-approx-2}, the scaling is $\eta \propto 1/\Omega_{\mathrm{eff}}^2$. Figure~\ref{fig:eta_vs_nu} shows $\eta_{\text{num}}$ and $|\eta_{\mathrm{approx}} - \eta_{\mathrm{num}}|/\eta_{\mathrm{num}}$ as functions of the trap frequency $\nu_{\mathrm{trap}}$ for various values of the atom temperature $\mathcal{T}$ and a fixed value of the Rabi frequency ($\Omega_{\mathrm{eff}}/2\pi = 300$~kHz). As expected from Eq.~\eqref{eq:eta-approx-2}, $\eta$ is independent of the trap frequency in the high-temperature regime ($\hbar \omega_{\mathrm{trap}} \ll k_{\mathrm{B}} \mathcal{T}$) and scales linearly with $\nu_{\mathrm{trap}}$ in the low-temperature regime ($\hbar \omega_{\mathrm{trap}} \gg k_{\mathrm{B}} \mathcal{T}$). We observe that the value of the relative error is very close to the value of $\eta_{\text{num}}$ and scales in exactly the same way. This is not surprising, since the leading term in $|\eta_{\mathrm{approx}} - \eta_{\mathrm{num}}|$ is on the order of $\braket{r^4}$ and hence $|\eta_{\mathrm{approx}} - \eta_{\mathrm{num}}|/\eta_{\mathrm{num}} \sim \braket{r^4}/\braket{r^2} \sim \braket{r^2} \sim \eta_{\mathrm{num}}$. 

Figure~\ref{fig:Dphi_vs_N} shows the phase uncertainty $(\Delta\phi)_{\mathrm{num}}$ and the relative error $|(\Delta\phi)_{\mathrm{approx}} - (\Delta\phi)_{\mathrm{num}}|/(\Delta\phi)_{\mathrm{num}}$ as functions of the number of entangled atoms, $N$, for three different parameter combinations that are described in Table~\ref{table:curve_parameters}. These three parameter combinations correspond to $\eta \approx 5.0 \times 10^{-3}$ (yellow curve), $\eta \approx 2.0 \times 10^{-3}$ (red curve), and $\eta \approx 5.0 \times 10^{-4}$ (blue curve). Figure~\ref{fig:Dphi_vs_N}(b) demonstrates that the difference between the numeric and approximate values of $\Delta\phi$ is comfortably small for all considered values of $N$. 

\begin{table}[htbp]
\caption{Parameter values for curves in Fig.~\ref{fig:Dphi_vs_N}.}
\begin{center}
\bgroup
\def\arraystretch{1.3}
\begin{tabular}{lrrrrr}
\hline\hline
Curve & $\Omega_{\mathrm{eff}} / 2\pi$ & $\nu_{\mathrm{trap}}$ & \multicolumn{1}{c}{$\mathcal{T}$} &  \multicolumn{1}{c}{$\eta$} & $N_{\ast}$ \\
 & \multicolumn{1}{c}{[kHz]} & [kHz] & [$\mu$K] & & \\
\hline
yellow & $\ \ \ 400$ & $\ \ \ \ 14.5$ & $\ \ \ \ 0.65$ & $\ \ \ \ 5.0 \times 10^{-3}$ & $\ \ \ \ 20$ \\
red & $\ \ \ 450$ & $\ \ \ \ 10.0$ & $\ \ \ \ 0.30$ & $\ \ \ \ 2.0 \times 10^{-3}$ & $\ \ \ \ 52$ \\
blue & $\ \ \ 600$ & $\ \ \ \ 5.9$ & $\ \ \ \ 0.10$ & $\ \ \ \ 5.0 \times 10^{-4}$ & $\ \ \ \ 209$ \\
\hline\hline
\end{tabular}
\egroup
\end{center}
\label{table:curve_parameters}
\end{table}%
  
To quantify how close a phase uncertainty curve is to the HL, we use the number $N_{\ast}$, which is defined as the number of atoms, for which the deviation from the HL is less than 10\%, i.e., $N_{\ast}$ is the largest $N$ for which $(1 - \eta)^N \geq 0.9$. The values of $N_{\ast}$ for the curves in Fig.~\ref{fig:Dphi_vs_N} are reported in Table~\ref{table:curve_parameters}. We see that, with optimistic parameter values, the deviation from the HL is small for $N \sim 100$. 

Note that the relative ac Stark shift $\delta^{\mathrm{AC}}$ in Eq.~\eqref{eq:delta12-at-resonance-simplified} has its own uncertainty, $\Delta \delta^{\mathrm{AC}}$, which arises due to laser intensity fluctuations. Now we can justify why we neglected this uncertainty in the analysis above. This uncertainty can be upper bounded by using the fact that $|\delta^{\mathrm{AC}}| \leq \Omega_{\text{eff}}$. Therefore, for a pulse of duration $\tau$ and area $A = \Omega_{\text{eff}} \tau$, we obtain:
\begin{equation}
(\Delta \delta^{\mathrm{AC}})^2 \leq (\Delta \Omega_{\text{eff}})^2 = \frac{\sigma_A^2}{\tau^2} = \frac{\xi^2 \Omega_{\text{eff}}^2 }{A} ,
\end{equation}
where $\sigma_A^2$ is the variance of the probability distribution for the error in $A$, and $\xi^2$ is the proportionality factor in the relationship $\sigma_A^2 = \xi^2 A$, as described in Sec.~\ref{sec:intensity-fluctuations}. Since $\xi \lesssim 10^{-3}$, the ratio
\begin{equation}
\frac{ (\Delta \delta^{\mathrm{AC}})^2 }{ (\Delta \delta'_{12})_{\text{th}}^2 } \leq \frac{\xi^2}{A \braket{r^2}} \approx \frac{\xi^2}{\eta} 
\label{eq:uncertainty-ratio-for-ac-Stark-shift}
\end{equation}
is very small. Even for $\eta \sim 10^{-4}$, which is smaller than any of the values in Table~\ref{table:curve_parameters}, the ratio in Eq.~\eqref{eq:uncertainty-ratio-for-ac-Stark-shift} is upper bounded by $\sim 10^{-2}$. Therefore, the uncertainty in the relative ac Stark shift due to laser intensity fluctuations can be safely neglected compared to the uncertainty in the Doppler shift term due to the initial momentum spread of the atoms.

\section{Effect of measurement error}
\label{sec:parity-measurement}

\subsection{Parity measurement protocol}

The analysis of the AI performance in this paper is based on the assumption that the expectation value of the parity operator $\Pi$ of Eq.~\eqref{eq:Parity-operator} is measured for a system of $N$ atoms. Therefore, an important question is how the number of measurements required to evaluate $\braket{\Pi}$ scales with $N$. We present here a protocol for parity measurement that scales linearly with $N$ despite the fact that the Hilbert space dimension grows exponentially.

The Hilbert space of a system of $N$ two-level atoms is spanned by the set $\mathfrak{S}$ of $2^N$ basis states, 
\begin{equation}
\mathfrak{S} = \{ |j\rangle | j = 0,1,\ldots,2^N - 1\}, 
\end{equation}
which are defined in Table~\ref{table:basis_states}.
\begin{table}[htbp]
\caption{Basis states for a system of $N$ two-level atoms.}
\begin{center}
\bgroup
\def\arraystretch{1.3}
\begin{tabular}{p{45mm}cc}
\hline\hline
$|j\rangle$ & $M_j$ & $\langle j | \Pi |j \rangle$ \\
\hline
$|0\rangle = |g \ldots g\rangle$ & $0$ & $+1$ \\
\hline
$|1\rangle = |g g\ldots ge\rangle$ & $1$ & $-1$ \\
$|2\rangle = |g g\ldots eg\rangle$ & $1$ & $-1$ \\
$\vdots$ & $\vdots$ & $\vdots$ \\
$|N\rangle = |e g\ldots gg\rangle$ & $1$ & $-1$ \\
\hline
$|N+1\rangle = |ggg\ldots gee\rangle$ & $2$ & $+1$ \\
$|N+2\rangle = |ggg\ldots ege\rangle$ & $2$ & $+1$ \\
$\vdots$ & $\vdots$ & $\vdots$ \\
$|N (N+1)/2\rangle = |eeg\ldots ggg\rangle$ & $2$ & $+1$ \\
\hline
$\vdots$ & $\vdots$ & $\vdots$ \\
\hline
$|2^N - 1\rangle = |e \ldots e\rangle$ & $N$ & $(-1)^N$ \\
\hline\hline
\end{tabular}
\egroup
\end{center}
\label{table:basis_states}
\end{table}%

Here, $M_j$ denotes the number of atoms in the $|e\rangle$ level for the state $|j\rangle$. From Table~\ref{table:basis_states}, it is easy to see that the entire set $\mathfrak{S}$ consists of $N+1$ subsets $\mathfrak{S}_M$ such that
\begin{equation}
\mathfrak{S}_M = \{ |j\rangle | M_j = M\}, \quad M = 0,1,\ldots, N ,
\end{equation}
i.e., all states $|j\rangle$ in $\mathfrak{S}_M$ have a fixed value of $M_j = M$ and, consequently, they all have the same expectation value $\langle j | \Pi |j \rangle = (-1)^M$. The subset $\mathfrak{S}_M$ includes $\binom{N}{M}$ states.

For an arbitrary state
\begin{equation}
|\psi\rangle = \sum_{j=0}^{2^N-1} c_j |j\rangle,
\end{equation}
the expectation value of the parity operator is
\begin{equation}
\braket{\Pi} = \sum_{j=0}^{2^N-1} | c_j |^2 \langle j | \Pi |j \rangle .
\end{equation}
Using the partition of the basis states $\ket{j}$ into the subsets $\{ \mathfrak{S}_M | M = 0,1,\ldots, N \}$, we obtain:
\begin{equation}
\label{eq:Parity}
\braket{\Pi} = \sum_{M = 0}^{N} P_M  (-1)^M  ,
\end{equation}
where
\begin{equation}
P_M = \sum_{|j\rangle \in \mathfrak{S}_M} | c_j |^2 
\end{equation}
is the probability that the system is in the subset $\mathfrak{S}_M$ of states.

Consider now a state-selective measurement, for example, via state-dependent fluorescence imaging (SFI)~\cite{Martinez.PRL.119.180503.2017, Kwon.PRL.119.180504.2017} or via coherent spatial splitting (CSS) in a state-dependent optical lattice~\cite{Wu.NatPhys.15.538.2019}, which detects whether an atom is in the $|e\rangle$ level or in the $|g\rangle$ level. For example, in SFI, by driving a resonant cycling transition, fluorescence is produced if the atom is in the $|e\rangle$ level and no fluorescence is produced if the atom is in the $|g\rangle$ level. Similarly, in CSS, by applying a sequence of pulses and lattice transformations, atoms in the $|e\rangle$ level will be shifted to the left and atoms in the $|g\rangle$ level will be shifted to the right. Hence, for a system of $N$ atoms, any state in the subset $\mathfrak{S}_M$ will produce $M$ fluorescence images (in SFI) or $M$ atoms shifted to the left (in CSS). Correspondingly, for an arbitrary state $|\psi\rangle$, the probability to detect $M$ fluorescence images (in SFI) or $M$ atoms shifted to the left (in CSS) is $P_M$. Therefore, the measurement of the $N+1$ probabilities $\{ P_M \}$ directly yields the parity expectation value via Eq.~(\ref{eq:Parity}).

\subsection{Measurement error}
\label{sec:ME}

For this measurement protocol, a conservative estimate of the effect of measurement error is based on assuming that a random wrong value of $\braket{\Pi}$ is obtained if at least one atom is detected in a wrong level ($|g\rangle$ instead of $|e\rangle$ or $|e\rangle$ instead of $|g\rangle$). By averaging over all these error outcomes, we obtain:
\begin{equation}
\overline{\braket{\Pi}} \approx (1-q_{\text{det}})^N \cos( N \phi ) ,
\label{eq:Parity-ME}
\end{equation}
where $q_{\text{det}}$ is the probability of erroneous state detection for one atom. Correspondingly, the phase uncertainty, minimized for the measurement at a dark fringe, where $\cos( N \phi ) = 0$, is
\begin{equation}
\label{Delta-phi-ME}
\Delta\phi \approx \frac{1}{(1-q_{\text{det}})^N N} .
\end{equation}
Equations~\eqref{eq:Parity-ME} and \eqref{Delta-phi-ME} comply, respectively, with the general forms $\overline{\braket{\Pi}} = \Pi_0 \cos( N \phi )$ and $\Delta\phi = (\Pi_0 N)^{-1}$, with $\Pi_0 \approx (1-q_{\text{det}})^N$. A further approximation $\Pi_0 \approx 1 - N q_{\text{det}}$ holds when $N q_{\text{det}} \ll 1$.

\begin{figure}[htbp]
	\centering
	\includegraphics[width=1.0\linewidth]{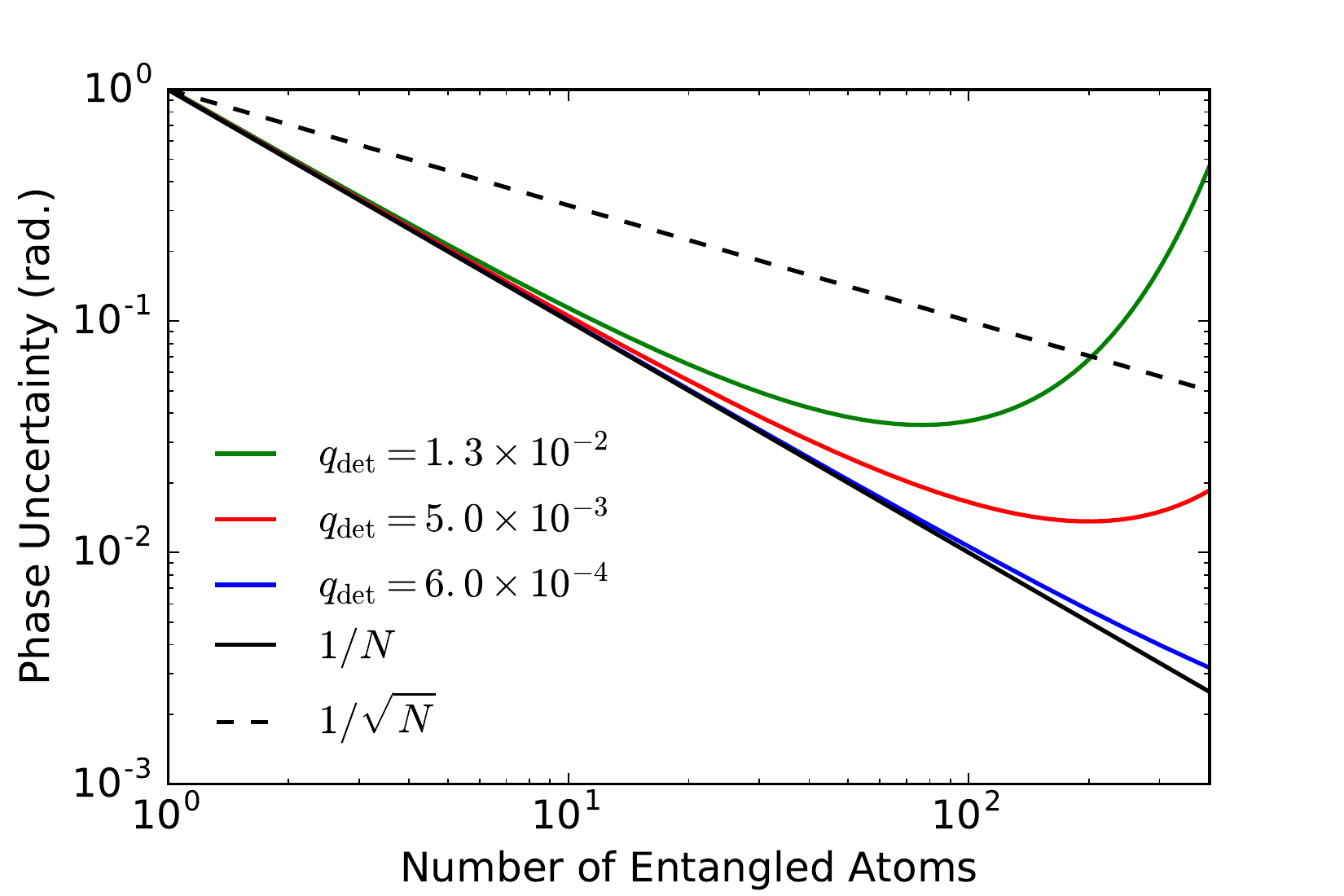}
	\caption{The phase uncertainty $\Delta\phi$ of Eq.~\eqref{Delta-phi-ME} as a function of the number of entangled atoms in the GHZ state, $N$, for various values of the state detection error probability for one atom, $q_{\text{det}}$. The HL $\Delta\phi = 1/N$ and the SQL $\Delta\phi = 1/N^{1/2}$ are also shown for comparison.}
	\label{fig:Figure_Phase_Uncertainty_Measurement_Error}
\end{figure}

In an SFI experiment with 10 atoms, an average state detection fidelity of 0.987 was reported~\cite{Kwon.PRL.119.180504.2017}, which corresponds to $(1-q_{\text{det}})^N \approx 0.877$ for $N = 10$. While fidelity scaling to a much larger $N$ is not yet known in SFI experiments, the same fidelity for $N = 100$ would result in $(1-q_{\text{det}})^N \approx 0.27$.
In a CSS experiment with 160 atoms, an average state detection fidelity of 0.9994 was reported~\cite{Wu.NatPhys.15.538.2019}, which corresponds to $(1-q_{\text{det}})^N \approx 0.94$ for $N = 100$. Furthermore, in CSS, fidelity is essentially independent of the number of atoms measured, and the measurement causes negligible atom loss. However, in order to perform a CSS experiment, atoms need to be loaded into a three-dimensional optical lattice.

Figure~\ref{fig:Figure_Phase_Uncertainty_Measurement_Error} shows the phase uncertainty $\Delta\phi$ of Eq.~\eqref{Delta-phi-ME} as a function of the number of entangled atoms in the GHZ state, $N$, for $q_{\text{det}} = 1.3 \times 10^{-2}$ (corresponding to the average state detection fidelity of 0.987 reported in the SFI experiment~\cite{Kwon.PRL.119.180504.2017}), $q_{\text{det}} = 5.0 \times 10^{-3}$ (a hypothetical value), and $q_{\text{det}} = 6.0 \times 10^{-4}$ (corresponding to the average state detection fidelity of 0.9994 reported in the CSS experiment~\cite{Wu.NatPhys.15.538.2019}). If the state detection error is sufficiently small (one per cent or lower), the phase uncertainty surpasses the SQL for $N \sim 100$.

\section{Effect of spontaneous emission}

In deriving the effective two-level model of Eq.~\eqref{eq:Ham-eff}, spontaneous emission from the intermediate level $\ket{i}$ has been neglected. This process (which is also referred to as spontaneous photon scattering) has been thoroughly analyzed in the context of high-fidelity quantum operations on trapped-ion hyperfine qubits~\cite{Wineland.PTRSLA.361.1349.2003, Ozeri.PRA.75.042329.2007, Ballance.PRL.117.060504.2016}. In general, the spontaneous emission rate, $R_{\text{SE}}$, depends on intensities of the Raman fields, detunings from the P levels, and their decay rates (natural line widths). In AI experiments with $^{133}$Cs atoms, the detuning from the $6^2\text{P}_{3/2}$ level, $\Delta/2\pi$, is typically in the range of $20$--$200$~GHz, while the fine structure splitting (the separation between the $6^2\text{P}_{1/2}$ and $6^2\text{P}_{3/2}$ levels), is $\omega_\text{F}/2\pi = 16.61$~THz. In this regime ($\Delta \ll \omega_\text{F}$), we can safely neglect the contribution to spontaneous emission from the $6^2\text{P}_{1/2}$ level, and obtain:
\begin{equation}
R_{\text{SE}} \approx \gamma_i P_i \approx \frac{\gamma_i \Omega_{\text{eff}}}{2 \Delta} ,
\end{equation}
where $P_i$ is the probability that the intermediate state $\ket{i} = \ket{6^2\text{P}_{3/2}}$ is occupied, and $\gamma_i /2\pi = 5.234$~MHz is its decay rate. For a pulse of duration $\tau$ and area $A = \Omega_{\text{eff}} \tau$, the probability of spontaneous emission is
\begin{equation}
P_{\text{SE}}(A) = R_{\text{SE}} \tau \approx \frac{A}{2} \frac{\gamma_i}{\Delta} .
\end{equation}

Similarly to the case of measurement errors discussed in Sec.~\ref{sec:ME} above, a conservative estimate of the effect of spontaneous emission is based on assuming that a random wrong value of $\braket{\Pi}$ is obtained if at least one atom undergoes spontaneous emission from the intermediate level. By averaging over all these erroneous outcomes, we obtain: $\overline{\braket{\Pi}} = \Pi_0 \cos( N \phi )$ and $\Delta\phi = (\Pi_0 N)^{-1}$, with
\begin{equation}
\Pi_0 \approx (1-q_{\text{SE}})^N , 
\end{equation}
where
\begin{equation}
q_{\text{SE}} \approx P_{\text{SE}}(\pi) + P_{\text{SE}}(\pi/2) \approx \frac{3 \pi}{4} \frac{\gamma_i}{\Delta} 
\end{equation}
is the probability of spontaneous emission for one atom in one AI experiment. A further approximation $\Pi_0 \approx 1 - N q_{\text{SE}}$ holds when $N q_{\text{SE}} \ll 1$. With $\Delta/2\pi$ ranging from $20$~GHz to $200$~GHz for $^{133}$Cs atoms, we find $q_{\text{SE}}$ ranging from $6.2 \times 10^{-4}$ to $6.2 \times 10^{-5}$, respectively. The deviation from the HL due to spontaneous emission would be less that 10\% (i.e., $\Pi_0 \geq 0.9$) for $N \leq 170$ and $N \leq 1708$, respectively.

\section{Effect of atom loss}

The loss of an atom during the AI operation can be detected, and the outcomes of the respective experiments can be eliminated from the data. Hence, the possibility of losing an atom leads to a reduced data-acquisition rate. At the end of our AI protocol, each atom is imaged individually by SFI~\cite{Martinez.PRL.119.180503.2017, Kwon.PRL.119.180504.2017}, and  the number of atoms in the bright state is recorded. Atom loss is then detected by transferring the dark-state atoms to the bright state, repeating the SFI measurement, and comparing the total number of atoms detected with the initial number of atoms. Therefore, post-selecting for the measurement outcomes without atom loss eliminates the error introduced by counting a lost atom as a dark-state measurement. 

This method of post-selection of lossless outcomes is not always possible in other systems. For example, in optical interferometry the loss of a single photon transforms the maximally entangled NOON state into an incoherent mixture \cite{Escher_2011, Pezze.RMP.90.035005.2018}. Without the ability to discriminate between lossless and lossy operations, the outcomes of measurements corresponding to incoherent mixture states add to the phase uncertainty of an optical interferometer.

The effect of a reduced data-acquisition rate due to atom loss on the phase uncertainty can be quantified as follows. If an AI experiment with the $N$-atom GHZ state is repeated $M$ times, then, in the ideal case, the phase uncertainty after $M$ experiments is
\begin{equation}
(\Delta \phi)_M = \frac{\Delta \phi}{\sqrt{M}} = \frac{1}{\sqrt{M} N} .
\end{equation}
If the loss probability for one atom in one AI experiment is $q_{\text{loss}}$, then the number of post-selected lossless outcomes is $M (1-q_{\text{loss}})^N$. The resulting phase uncertainty after $M$ experiments is
\begin{equation}
(\Delta \phi)_M = \frac{1}{\sqrt{M (1-q_{\text{loss}})^N} N} .
\label{eq:Delta-phi-AL-M-cycles}
\end{equation}
From Eq.~\eqref{eq:Delta-phi-AL-M-cycles}, the scaling of the phase uncertainty per one experiment, $\Delta \phi$, with $N$ can be interpreted differently depending on the type of application, for which the AI is used. For example, in a stationary gravity measurement, it might be possible to repeat the experiment as many times as needed to achieve the desired number of lossless outcomes. This can be interpreted as if the atom loss prolonged the effective time that it takes to perform one lossless experiment but did not affect the scaling of $\Delta \phi$. However, in many cases (in particular, when the AI is used in an inertial navigation system), the total measurement time and hence the total number of experiments (AI cycles) are fixed. In such a case the result of Eq.~\eqref{eq:Delta-phi-AL-M-cycles} can be effectively interpreted as the phase uncertainty for $M$ lossless experiments with the uncertainty per one experiment given by
\begin{equation}
\Delta \phi = \frac{1}{(1-q_{\text{loss}})^{N/2} N} .
\label{Delta-phi-AL}
\end{equation}
This result has the general form $\Delta\phi = (\Pi_0 N)^{-1}$, where $\Pi_0 = (1-q_{\text{loss}})^{N/2}$. A further approximation $\Pi_0 \approx 1 - N q_{\text{loss}}/2$ holds when $N q_{\text{loss}}/2 \ll 1$. For example, with $2\%$ atom loss, we obtain $(1-q_{\text{loss}})^{N/2} \approx 0.90$ for $N = 10$ and $(1-q_{\text{loss}})^{N/2} \approx 0.36$ for $N = 100$. 

\begin{figure}[htbp]
	\centering
	\includegraphics[width=1.0\linewidth]{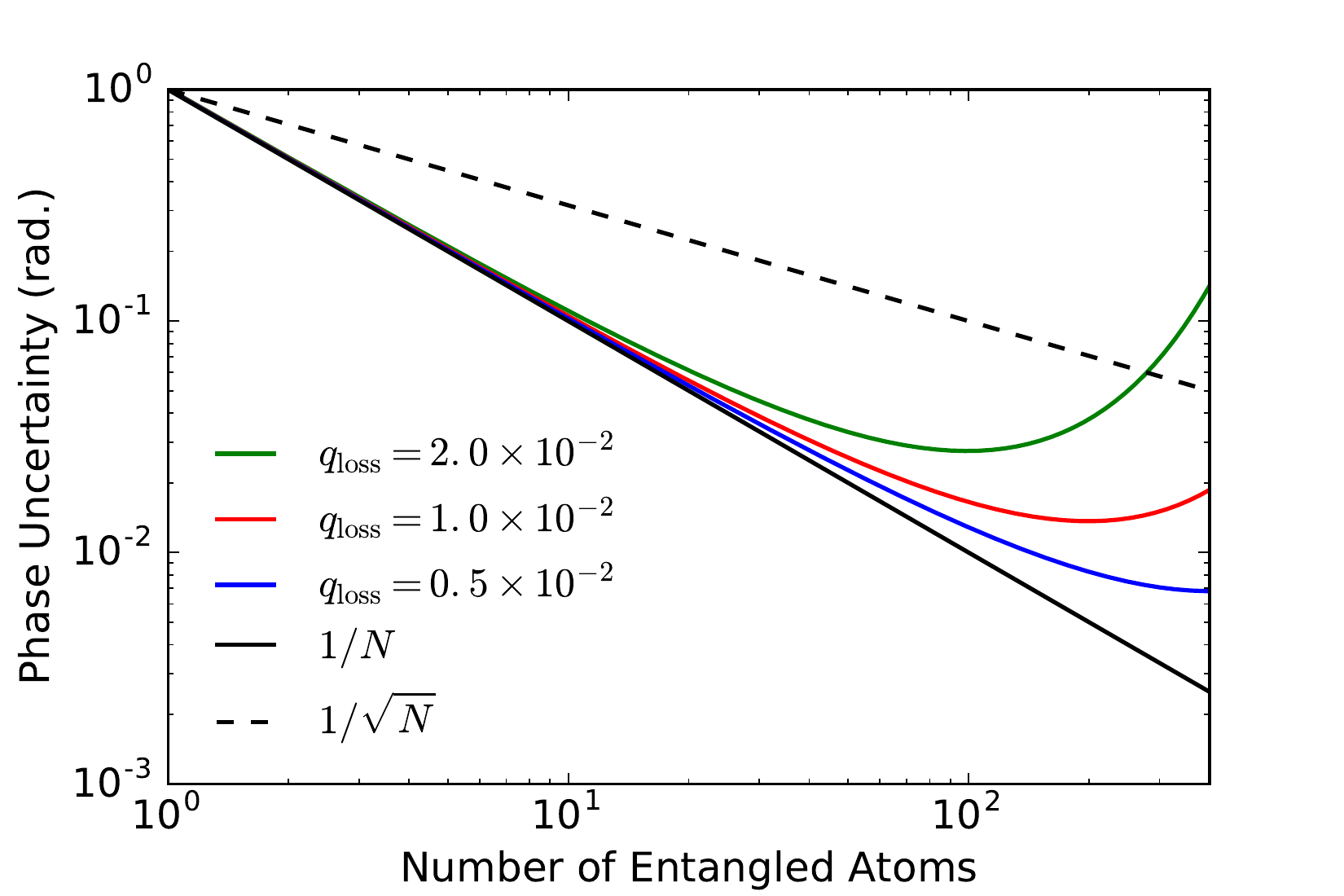}
	\caption{The phase uncertainty $\Delta\phi$ of Eq.~\eqref{Delta-phi-AL} as a function of the number of entangled atoms in the GHZ state, $N$, for various values of the loss probability for one atom in one AI experiment, $q_{\text{loss}}$. The HL $\Delta\phi = 1/N$ and the SQL $\Delta\phi = 1/N^{1/2}$ are also shown for comparison.}
	\label{fig:Figure_Phase_Uncertainty_Atom_Loss}
\end{figure}

Figure~\ref{fig:Figure_Phase_Uncertainty_Atom_Loss} shows the phase uncertainty $\Delta\phi$ of Eq.~\eqref{Delta-phi-AL} as a function of the number of entangled atoms in the GHZ state, $N$, for $q_{\text{loss}} = 2.0 \times 10^{-2}$, $q_{\text{loss}} = 1.0 \times 10^{-2}$, and $q_{\text{loss}} = 0.5 \times 10^{-2}$. We see that if atom loss is sufficiently low (two per cent or smaller), the phase uncertainty surpasses the SQL for $N \sim 100$.

\begin{table*}[htbp]
\caption{Summary of error sources that affect interferometry with entangled atoms. Each error source results in a reduction of the parity oscillation amplitude, $\Pi_0$, below the ideal value of $\Pi_0 = 1$. $N$ is the number of atoms in the initial GHZ state. $f_{\mathrm{WN}}(\theta;0,\sigma^2)$ is the probability density function of the wrapped normal distribution for variable $\theta$, with zero mean and variance $\sigma^2$.}
\begin{center}
\bgroup
\def\arraystretch{1.7}
\begin{tabular}{p{20mm}p{28mm}p{26mm}p{97mm}}
\hline
\hline
 Error source & 
 $\Pi_0$ & 
 Approximate $\Pi_0$ \newline for small error  & 
 Error parameters \\
\hline
Initial state \newline preparation & $(1-q_{\zeta}) e^{-\sigma_{\beta}^2 /2}$ & $1 - (q_{\zeta}+\sigma_\beta^2/2)$ & 
$q_{\zeta}$ --- probability of admixed noise state $\ket{\zeta}\bra{\zeta}$ \newline 
$\beta$ --- random phase between two components of the initial GHZ state, \newline 
\hspace*{6mm} $P(\beta) = f_{\mathrm{WN}}(\beta;0,\sigma_{\beta}^2)$ \newline
$\Pi_0 = 2 \overline{\mathcal{F}} - 1$, where $\overline{\mathcal{F}}$ is the average fidelity of the initial GHZ state  
\\
\hline
Laser intensity \newline fluctuations & $(1 + N \xi^2 \pi/2)^{-1}$ \newline (valid for $N \gg 1$) & $1 - N \xi^2 \pi/2$ &
$v$ --- error in the $\pi/2$ pulse, $P(v) = f_{\mathrm{WN}}(v;0,\sigma_v^2)$ \newline
$w$ --- error in the $\pi$ pulse, $P(w) = f_{\mathrm{WN}}(w;0,\sigma_w^2)$ \newline
$\xi^2$ ---  proportionality factor, $\sigma_v^2 = \xi^2 \pi/2$, $\sigma_w^2 = \xi^2 \pi$ \\
\hline
Laser phase \newline fluctuations & $\exp(- N^2 \sigma_{\vartheta}^2 r_{\text{corr}}/2)$ & $1-N^2\sigma_\vartheta^2 r_{\text{corr}}/2$ & 
$\vartheta_{\pi}$ --- random phase in the $\pi$ pulse, $P(\vartheta_{\pi}) = f_{\mathrm{WN}}(\vartheta_{\pi};0,\sigma_{\vartheta}^2)$ \newline
$\vartheta_{\pi/2}$ --- random phase in the $\pi/2$ pulse, $P(\vartheta_{\pi/2}) = f_{\mathrm{WN}}(\vartheta_{\pi/2};0,\sigma_{\vartheta}^2)$ \newline
$\tilde{\vartheta}$ --- effective random phase, $\tilde{\vartheta} = 2 \vartheta_{\pi} - \vartheta_{\pi/2}$, 
$P(\tilde{\vartheta}) = f_{\mathrm{WN}}(\tilde{\vartheta};0, r_{\text{corr}} \sigma_{\vartheta}^2)$ \newline
$r_{\text{corr}}$ --- correlation factor, $1 \leq r_{\text{corr}} \leq 5$ \\
\hline
Initial \newline momentum \newline spread & $(1-\eta)^N$ & $1-N \eta$ & 
$\eta$ --- parameter that quantifies the error due to the initial momentum spread \newline
\hspace*{5.8mm}  of trapped atoms; the exact expression for $\eta$ is given by Eq.~\eqref{eq:eta-general} and \newline
\hspace*{5.8mm}  an approximate expression (valid for small errors) is given by Eq.~\eqref{eq:eta-approx-2} \\
\hline
Measurement & $(1-q_{\text{det}})^N$ & $1 - N q_{\text{det}}$ & 
$q_{\text{det}}$ --- probability of erroneous state detection for one atom in a \newline 
\hspace*{8.1mm} state-selective measurement \\
\hline
Spontaneous \newline emission & $(1-q_{\text{SE}})^N$ & $1 - N q_{\text{SE}}$ & 
$q_{\text{SE}}$ --- probability of spontaneous emission for one atom in one AI \newline 
\hspace*{8.1mm} experiment; $q_{\text{SE}} \approx (3 \pi /4) \gamma_i / \Delta$, where $\Delta$ is the detuning from \newline 
\hspace*{8.1mm} the intermediate level $\ket{i}$ and $\gamma_i$ is its decay rate (natural line width)\\
\hline
Atom loss & $(1-q_{\text{loss}})^{N/2}$ & $1 - N q_{\text{loss}}/2$ & 
$q_{\text{loss}}$ --- loss probability for one atom in one AI experiment \\
\hline
\hline
\end{tabular}
\egroup
\end{center}
\label{tab:errors}
\end{table*}%

\section{Summary and conclusions}
\label{sec:conc}

Within the protocol for AI operation that we have described, non-negligible errors come from the imperfect initial state preparation, intensity and phase fluctuations of the Raman beams, the initial momentum spread of the atoms in the optical traps, imperfect measurement, spontaneous emission during Raman pulses, and atom loss. A key finding is that each of these errors preserves the general forms for the parity oscillations, $\overline{\braket{\Pi}} = \Pi_0 \cos( N \phi )$, and for the minimum phase uncertainty, $\Delta\phi = (\Pi_0 N)^{-1}$, and manifests itself through a reduction of the amplitude $\Pi_0$ below the ideal value of $\Pi_0 = 1$. We have derived analytical results that express the dependence of $\Pi_0$ on error parameters and $N$ for all of these error sources. When an error is small, $\Pi_0$ is close to $1$ and can be approximated as $\Pi_0 \approx 1 - \varepsilon$, where $\varepsilon \ll 1$.

If we use an index $\ell$ to enumerate different uncorrelated error sources, then the total effect of all errors on the amplitude of parity oscillations is given by
\begin{equation}
\Pi_0^{\mathrm{total}} = \prod_{\ell} \Pi_0^{(\ell)} .
\end{equation}
When all of these errors are small ($\varepsilon_{\ell} \ll 1 \ \forall \ell$), the total error is simply a sum of all the individual errors: $\varepsilon_{\mathrm{total}} \approx \sum_{\ell} \varepsilon_{\ell}$.

Table \ref{tab:errors} reports a summary of all the non-negligible error sources considered in this work. This table lists exact and approximate (valid for small error) expressions for $\Pi_0$ for each error source, along with descriptions of respective error parameters. Currently, error in the initial state preparation is not prohibitive for $N \lesssim 20$. For this error, an explicit scaling with $N$ is not available, and generating larger high-fidelity entangled states is an active field of study. Errors arising from intensity and phase fluctuations of the Raman beams are insignificant for large numbers of atoms ($N \lesssim 10^4$), assuming state-of-the-art optical technology. The error due to the initial momentum distribution of the atoms scales linearly with the average vibrational energy of the atom in the trap and inversely proportional to the square of the effective Rabi frequency for the two-photon Raman transition. For optimistic (yet still realistic) parameter values (atoms cooled to sub-microkelvin temperatures in shallow traps and driven with very intense Raman fields), this error can be made insignificant for $N \lesssim 100$. Optimal shaping of Raman fields is potentially useful for mitigating this error~\cite{Saywell.PRA.98.023625.2018, Saywell_2020}. Measurement error and the effect of atom loss can also be significant for large numbers of atoms, but attaining a phase uncertainty near the HL still appears to be feasible for $N \sim 100$. The CSS method of state-selective measurement, while challenging in an AI setup, promises superior scaling with the number of atoms~\cite{Wu.NatPhys.15.538.2019}.

The obtained results indicate that an entanglement-enhanced AI with a phase uncertainty close to the HL is feasible for $N \sim 100$ with state-of-the-art experimental capabilities. This would be equivalent to a phase uncertainty of an AI with $\sim 10^4$ independent atoms, operated at the SQL. Therefore, it is important to understand which technology advances are needed to make interferometry with entangled atoms preferable over the possibility of simply increasing the number of independent atoms. Based on the presented analysis, achieving a phase uncertainty close to the HL for $N \sim 1000$ would require significant improvements in four key areas: (1) generation of the GHZ state with a high fidelity ($\overline{\mathcal{F}} \gtrsim 0.95$) for such a large $N$, (2) better atom cooling, improvements in laser intensity and focusing, and optimal shaping of Raman fields to decrease the momentum spread error to the level of $\eta \lesssim 10^{-4}$, (3) improvements in state-selective measurements (e.g., via further refinement of the CSS method) to reduce the error probability to the level of $q_{\text{det}} \lesssim 10^{-4}$, and (4) improvements in atom manipulation and control to reduce the atom loss probability to the level of $q_{\text{loss}} \lesssim 2 \times 10^{-4}$. With these potential advancements, an entanglement-enhanced AI could achieve a phase uncertainty equivalent to that of an AI with $\sim 10^6$ independent atoms, which is near the limit posed by miniaturization requirements for some applications.

Interferometers with highly entangled atomic spin states show great promise as inertial sensors, and an array of single-atom optical tweezers provides a suitable platform for these sensors. When the atoms are cooled to ultracold temperatures and initiated in the GHZ state, only a few error sources stand in the way of reaching a phase uncertainty near the HL. We have shown how these errors scale with the number of entangled atoms and identified parameters that quantify these errors for a realistic system of entangled Cs atoms.
 
\acknowledgments
The authors acknowledge useful discussions with and suggestions from Jonathan Bainbridge, Matthew Chow, Yuan-Yu Jau, Shanalyn Kemme, Jongmin Lee, Bethany Little, Alberto Marino, Michael Martin, Lambert Paul Parazzoli, Mohan Sarovar, and Peter Schwindt.
This work is supported by the Laboratory Directed Research and Development program at Sandia National Laboratories. 
Sandia National Laboratories is a multimission laboratory managed and operated by National Technology and Engineering Solutions of Sandia, LLC, a wholly owned subsidiary of Honeywell International, Inc., for the U.S. Department of Energy's National Nuclear Security Administration under contract DE-NA0003525. This paper describes objective technical results and analysis. Any subjective views or opinions that might be expressed in the paper do not necessarily represent the views of the U.S. Department of Energy or the United States Government.

\appendix
\section{Derivation of the parity expectation value for various initial states and pulse sequences}
\label{sec:app-derivation}

Consider the evolution operator $U_{\text{tot}}^{(k)}$ for an entire sequence of the AI pulses acting on the $k$th atom, in the basis $\{ |g, \mathbf{p}_k \rangle_k, |e, \mathbf{p}_k + \hbar\mathbf{K} \rangle_k \}$. Note that we omitted the time dependence from the basis states, since, according to Eq.~\eqref{eq:Parity-EV-general}, the field-free evolution has no effect on the parity expectation value. In this basis, $U_{\text{tot}}^{(k)}$ has the general matrix form:
\begin{equation}
U_{\text{tot}}^{(k)} = \begin{bmatrix}
U_{gg}^{(k)} & U_{ge}^{(k)} \\[0.3em] U_{eg}^{(k)} & U_{ee}^{(k)}
\end{bmatrix} . 
\label{eq:app:U-tot}
\end{equation}
In general, the matrix elements in Eq.~\eqref{eq:app:U-tot} depend on the phases $\phi_{T}^{(k)}$ and $\phi_{2T}^{(k)}$ which, in their turn, depend on the momentum $\mathbf{p}_k$ through the Doppler term $-\mathbf{p}_k \cdot \mathbf{K} / m$ in the Raman detuning $\delta_{12}^{(k)}$ experienced by the $k$th atom. Following Eq.~\eqref{eq:U}, the evolution operator for the system of $N$ atoms is given by the tensor product of all one-atom evolution operators: $U_{\text{tot}} = \bigotimes_{k=1}^N U_{\text{tot}}^{(k)}$.

First consider an initial state for $N$ atoms which are not entangled. The general form for an unentangled state is
\begin{equation}
|\Psi\rangle_{\text{in}} = \bigotimes_{k=1}^N \left[ c_g^{(k)} |g, \bar{\mathbf{p}}_k \rangle_k 
+ c_e^{(k)} |e, \bar{\mathbf{p}}_k + \hbar\mathbf{K} \rangle_k \right] ,
\label{eq:app:state-product-in}
\end{equation}
where $c_g^{(k)}$ and $c_e^{(k)}$ are complex coefficients (subject to the normalization condition $|c_g^{(k)}|^2 + |c_e^{(k)}|^2 = 1$), and each of the component states is a linear superposition of the basis states:
\begin{equation}
|\alpha,\bar{\mathbf{p}}_k+\hbar \mathbf{K}_{\alpha}\rangle_k 
= \int d^3 p_k\, \tilde{\psi}_{\bar{\mathbf{p}}_k}(\mathbf{p}_k)  |\alpha,\mathbf{p}_k+\hbar \mathbf{K}_{\alpha}\rangle_k ,
\end{equation}
for $\alpha = \{g,e\}$ with $\mathbf{K}_{g} = 0$ and $\mathbf{K}_{e} = \mathbf{K}$. The output state $|\Psi\rangle_{\text{out}} = U_{\text{tot}} |\Psi\rangle_{\text{in}}$ is given by
\begin{align}
|\Psi\rangle_{\text{out}} = & \bigotimes_{k=1}^N \int d^3 p_k\, \tilde{\psi}_{\bar{\mathbf{p}}_k}(\mathbf{p}_k) \nonumber \\
& \times \left[ \left( c_g^{(k)} U_{gg}^{(k)} + c_e^{(k)} U_{ge}^{(k)} \right) |g, \mathbf{p}_k \rangle_k \right. \nonumber \\
& + \left. \left( c_g^{(k)} U_{eg}^{(k)} + c_e^{(k)} U_{ee}^{(k)} \right) |e, \mathbf{p}_k + \hbar\mathbf{K} \rangle_k \right] .
\label{eq:app:state-product-out}
\end{align}
By substituting this result into Eq.~\eqref{eq:Parity-EV-general} and using the unitarity of the matrix $U_{\text{tot}}^{(k)}$, we obtain:
\begin{align}
\braket{\Pi} =\, & \prod_{k=1}^N \left[ 
2 \text{Re}\! \left( c_g^{(k) \ast} c_e^{(k)} \left\langle 2 U_{gg}^{(k) \ast} U_{ge}^{(k)} \right\rangle \right) \right. \nonumber \\
& + \left. \left( |c_g^{(k)}|^2 - |c_e^{(k)}|^2 \right) \left\langle |U_{gg}^{(k)}|^2 - |U_{eg}^{(k)}|^2 \right\rangle
 \right] ,
\label{eq:app:Parity-EV-product-1}
\end{align}
where we introduced the notation:
\begin{equation}
\left\langle f^{(k)} \right\rangle \equiv \int d^3 p_k\, |\tilde{\psi}_{\bar{\mathbf{p}}_k}(\mathbf{p}_k)|^2 f^{(k)}(\mathbf{p}_k) 
\end{equation}
for averaging over the momentum distribution of the $k$th atom.

For the three-pulse sequence $\pi/2$--$\pi$--$\pi/2$, the evolution operator for one atom is given by Eq.~\eqref{eq:U-tot-3-pulses} or, explicitly for the $k$th atom:
\begin{equation}
U_{\text{tot}}^{(k)} = \frac{1}{2} \begin{bmatrix}
- e^{i \phi_T^{(k)}} (1+ e^{i\phi_k}) & - e^{i \phi_T^{(k)}} (1 - e^{i\phi_k}) \\
e^{-i \phi_T^{(k)}} (1 - e^{-i\phi_k})& - e^{-i \phi_T^{(k)}} (1 + e^{-i\phi_k}) 
\end{bmatrix} ,
\label{eq:app:U-tot-3-pulses} 
\end{equation}
where
\begin{equation}
\phi_k \equiv \phi_{2T}^{(k)} - 2 \phi_{T}^{(k)} = -\mathbf{K} \cdot \mathbf{a}_k T^2 . 
\label{eq:app:phi-k}
\end{equation}
Using the matrix elements from Eq.~\eqref{eq:app:U-tot-3-pulses}, we obtain:
\begin{subequations}
\label{eq:app:U-terms-3-pulses}
\begin{align}
& 2 U_{gg}^{(k) \ast} U_{ge}^{(k)} = -i \sin\phi_k , \\
& |U_{gg}^{(k)}|^2 - |U_{eg}^{(k)}|^2 = \cos\phi_k .
\end{align}
\end{subequations}
Since these terms depend only on $\phi_k$, and $\phi_k$ is independent of $\mathbf{p}_k$, each of the averages in Eq.~\eqref{eq:app:Parity-EV-product-1} is equivalent to multiplication by 1, regardless of the specific forms of the momentum distributions $|\tilde{\psi}_{\bar{\mathbf{p}}_k}(\mathbf{p}_k)|^2$. Therefore, by substituting Eqs.~\eqref{eq:app:U-terms-3-pulses} into Eq.~\eqref{eq:app:Parity-EV-product-1}, we obtain:
\begin{align}
\braket{\Pi} = & \prod_{k=1}^N \left[ 2 \text{Im}\! \left( c_g^{(k) \ast} c_e^{(k)} \right) \sin\phi_k \right. \nonumber \\
& + \left. \left( |c_g^{(k)}|^2 - |c_e^{(k)}|^2 \right) \cos\phi_k \right] .
\label{eq:app:Parity-EV-product-2}
\end{align}
In particular, if all atoms are initially prepared in the ground state, $c_g^{(k)} = 1, \ c_e^{(k)} = 0, \ \forall k$, then Eq.~\eqref{eq:app:Parity-EV-product-2} yields
\begin{equation}
\braket{\Pi} = \prod_{k=1}^N  \cos\phi_k .
\label{eq:app:Parity-EV-product-3}
\end{equation}

For the two-pulse sequence $\pi$--$\pi/2$, the evolution operator acting on the $k$th atom is given by Eq.~\eqref{eq:U-tot-2-pulses}. Using the matrix elements from Eq.~\eqref{eq:app:U-tot-3-pulses}, we obtain:
\begin{subequations}
\label{eq:app:U-terms-2-pulses}
\begin{align}
& 2 U_{gg}^{(k) \ast} U_{ge}^{(k)} = e^{-i \phi_k} , \\
& |U_{gg}^{(k)}|^2 - |U_{eg}^{(k)}|^2 = 0 .
\end{align}
\end{subequations}
Once again, these terms are independent of $\mathbf{p}_k$, and each of the averages in Eq.~\eqref{eq:app:Parity-EV-product-1} is equivalent to multiplication by 1, which produces
\begin{equation}
\braket{\Pi} = \prod_{k=1}^N 2 \text{Re}\! \left( c_g^{(k) \ast} c_e^{(k)} e^{-i \phi_k} \right) .
\label{eq:app:Parity-EV-product-4}
\end{equation}
In particular, for the state $|\zeta\rangle$ of Eq.~\eqref{eq:zeta-state}, the coefficients are $c_g^{(k)} = \cos (\vartheta_k/2)$ and $c_e^{(k)} = e^{i\varphi_k} \sin (\vartheta_k/2)$. With these coefficients, Eq.~\eqref{eq:app:Parity-EV-product-4} yields
\begin{equation}
\braket{\Pi} = \prod_{k=1}^N  \sin\vartheta_k \cos(\phi_k - \varphi_k).
\label{eq:app:Parity-EV-product-5}
\end{equation}

Next, consider an initial entangled state of the form
\begin{equation}
|\Psi\rangle_{\text{in}} = 
c_g \bigotimes_{k=1}^{N}  |g,\bar{\mathbf{p}}_k \rangle_k + 
c_e \bigotimes_{k=1}^{N}  |e,\bar{\mathbf{p}}_k+\hbar \mathbf{K}\rangle_k ,
\label{eq:app:state-entangled-in}
\end{equation}
where $c_g$ and $c_e$ are complex coefficients (subject to the normalization condition $|c_g|^2 + |c_e|^2 = 1$). The output state $|\Psi\rangle_{\text{out}} = U_{\text{tot}} |\Psi\rangle_{\text{in}}$ is given by
\begin{align}
|\Psi\rangle_{\text{out}} =\, &
c_g \bigotimes_{k=1}^N \int d^3 p_k\, \tilde{\psi}_{\bar{\mathbf{p}}_k}(\mathbf{p}_k) \nonumber \\
& \times \left( U_{gg}^{(k)} |g, \mathbf{p}_k \rangle_k + U_{eg}^{(k)} |e, \mathbf{p}_k + \hbar\mathbf{K} \rangle_k \right) \nonumber \\
& + c_e \bigotimes_{k=1}^N \int d^3 p_k\, \tilde{\psi}_{\bar{\mathbf{p}}_k}(\mathbf{p}_k) \nonumber \\
& \times \left( U_{ge}^{(k)} |g, \mathbf{p}_k \rangle_k + U_{ee}^{(k)} |e, \mathbf{p}_k + \hbar\mathbf{K} \rangle_k \right) .
\label{eq:app:state-entangled-out}
\end{align}
By substituting this result into Eq.~\eqref{eq:Parity-EV-general} and using the unitarity of the matrix $U_{\text{tot}}^{(k)}$, we obtain:
\begin{align}
\braket{\Pi} = & 2 \text{Re}\! \left[ c_g^{\ast} c_e \prod_{k=1}^N \left\langle 2 U_{gg}^{(k) \ast} U_{ge}^{(k)} \right\rangle \right]  \nonumber \\
& + \left[ |c_g|^2 + (-1)^N |c_e|^2 \right] \prod_{k=1}^N \left\langle |U_{gg}^{(k)}|^2 - |U_{eg}^{(k)}|^2 \right\rangle .
\label{eq:app:Parity-EV-entangled-1}
\end{align}

For the two-pulse sequence $\pi$--$\pi/2$, we use the terms in Eqs.~\eqref{eq:app:U-terms-2-pulses} and substitute them into Eq.~\eqref{eq:app:Parity-EV-entangled-1} to obtain:
\begin{equation}
\braket{\Pi} = 2 \text{Re}\! \left[ c_g^{\ast} c_e \exp\left(-i \sum_{k=1}^N \phi_k \right) \right] .
\label{eq:app:Parity-EV-entangled-2}
\end{equation}
In particular, for the GHZ state of Eq.~\eqref{eq:GHZ-state-in}, $c_g = c_e = 1/\sqrt{2}$, which yields $\braket{\Pi} = \cos\left(\sum_{k=1}^N \phi_k \right)$. Similarly, for the state $| \Psi(\beta) \rangle$ of Eq.~\eqref{eq:Psi-beta-state}, $c_g = 1/\sqrt{2}$ and $c_e = e^{i\beta} /\sqrt{2}$, which yields $\braket{\Pi} = \cos\left(\sum_{k=1}^N \phi_k - \beta \right)$.

Next, we consider the two-pulse sequence $\pi$--$\pi/2$ in the presence of pulse-area errors due to laser intensity fluctuations. In this case, $U_{\text{tot}}^{(k)} = U_{2T}^{(k)} (\pi/2 + v) U_{T}^{(k)} (\pi + w)$, where $v$ and $w$ are the respective pulse-area errors. Using the matrix elements of $U_{\text{tot}}^{(k)}$, given by Eqs.~\eqref{eq:U-tot-2-pulses-noise}, we obtain:
\begin{subequations}
\label{eq:app:U-terms-2-pulses-noise}
\begin{gather}
2 U_{gg}^{(k) \ast} U_{ge}^{(k)} =  - \cos v \sin^2 (w/2) e^{i \phi_{2T}^{(k)}} - \sin v \sin w e^{i \phi_{T}^{(k)}} \nonumber \\
 + \cos v \cos^2 (w/2) e^{-i \phi_k} ,  \\
|U_{gg}^{(k)}|^2 - |U_{eg}^{(k)}|^2 = \cos v \sin w \cos\left( \phi_{2T}^{(k)} - \phi_{T}^{(k)} \right) \nonumber \\
 + \sin v \cos w  .
\end{gather}
\end{subequations}
Substituting these terms into Eq.~\eqref{eq:app:Parity-EV-entangled-1}, we encounter factors 
\begin{subequations}
\label{eq:app:av-all}
\begin{align}
\left\langle  e^{i \left( \phi_{2T}^{(k)} - \phi_{T}^{(k)} \right)} \right\rangle \propto &
\left\langle e^{-i \mathbf{p}_k \cdot \mathbf{K} T/m} \right\rangle , \label{eq:app:av-1} \\
\left\langle e^{i \phi_{T}^{(k)}} \right\rangle \propto &
\left\langle e^{-i \mathbf{p}_k \cdot \mathbf{K} T/m} \right\rangle , \label{eq:app:av-2} \\
\left\langle e^{i \phi_{2T}^{(k)}} \right\rangle \propto &
\left\langle e^{-2 i \mathbf{p}_k \cdot \mathbf{K} T/m} \right\rangle . \label{eq:app:av-3}
\end{align}
\end{subequations}
These averages are integrals over very rapidly oscillating functions and therefore they are extremely small. Specifically, for an atom in a harmonic trap, the averages in Eqs.~\eqref{eq:app:av-1} and \eqref{eq:app:av-2} scale as $e^{-\gamma}$ and the one in Eq.~\eqref{eq:app:av-3} scales as $e^{-4\gamma}$, where $\gamma = (KT \Delta p / m)^2$ and $\Delta p$ is the atom's momentum uncertainty. Even with the minimum uncertainty, $(\Delta p)_0 = \sqrt{\hbar m\omega_{\mathrm{trap}}/2}$, we find $\gamma \sim 10^{3}$ for typical values of $T$ ($\sim 1$~ms) and $\nu_{\mathrm{trap}}$ ($\sim 10$~kHz). After neglecting all terms that include these extremely small factors, we obtain:
\begin{align}
\braket{\Pi} =\, & 2 \cos^N\! v \cos^{2N} (w/2) \text{Re}\! \left[ c_g^{\ast} c_e e^{-i \sum_{k=1}^N \phi_k } \right] \nonumber \\
& + \left[ |c_g|^2 + (-1)^N |c_e|^2 \right]  \sin^N\! v \cos^N\! w .
\label{eq:app:Parity-EV-entangled-3}
\end{align}
Since the pulse-area error $v$ is small, $v \ll 1$, the second term in Eq.~\eqref{eq:app:Parity-EV-entangled-3} includes a factor that scales as $v^N$. This factor is extremely small in the regime of large atom numbers, $N \gg 1$, in which we are interested, and therefore the second term in Eq.~\eqref{eq:app:Parity-EV-entangled-3} can be safely neglected. Then we find:
\begin{equation}
\braket{\Pi} = 2 \cos^N\! v \cos^{2N} (w/2) \text{Re}\! \left[ c_g^{\ast} c_e e^{-i \sum_{k=1}^N \phi_k } \right] .
\label{eq:app:Parity-EV-entangled-4}
\end{equation}
In particular, for the GHZ state of Eq.~\eqref{eq:GHZ-state-in}, $c_g = c_e = 1/\sqrt{2}$, which yields
\begin{equation}
\braket{\Pi} = \cos^N\! v \cos^{2N} (w/2) \cos\left( \sum_{k=1}^N \phi_k \right) .
\label{eq:app:Parity-EV-entangled-5} 
\end{equation}

Finally, we consider the case where the pulse detuning error due to the initial momentum uncertainty of the atoms is taken into account, as described in Sec.~\ref{sec:Doppler-shift}. In this case, the initial state of the system is described by the density matrix $\rho(0)$ of Eq.~\eqref{eq:rho_intial_2}, and the parity expectation value is given by Eq.~\eqref{eq:Parity-EV-general-rho-1}. With the coefficients $X_{\vec{\alpha}} = c_g$ if $\alpha_k = g$ for all $k$, $X_{\vec{\alpha}} = c_e$ if $\alpha_k = e$ for all $k$, and $X_{\vec{\alpha}} = 0$ otherwise, Eq.~\eqref{eq:Parity-EV-general-rho-1} takes the form of Eq.~\eqref{eq:app:Parity-EV-entangled-1}, but now the average over the momentum distribution of the $k$th atom is
\begin{equation}
\left\langle f^{(k)} \right\rangle = \int_{-\infty}^{\infty} d p_k  P(p_k) f^{(k)}(p_k) ,
\label{eq:average-P}
\end{equation}
where $P(p_k) = \bra{p_k} \rho_{\mathrm{vib}}^{(k)} \ket{p_k}$ is the momentum distribution for the thermal vibrational state of one atom, given by Eq.~\eqref{eq:S-final}. Since this momentum distribution is the same for all atoms, in what follows we will rename the integration variable from $p_k$ to $p$ in the averages of the form~\eqref{eq:average-P}, to simplify the notation.

When the momentum spread is taken into account, the evolution operator for the two-pulse sequence $\pi$--$\pi/2$ is given by Eq.~\eqref{eq:U-2-pulses-with-detunerr}, and the evolution operator for each of the pulses in Eq.~\eqref{eq:U-2-pulses-with-detunerr}, $U_{t}^{(k)}(\tau)$, is given by Eq.~\eqref{eq:U-general}, where $t = T$, $\tau = \tau_{\pi}$ for the $\pi$ pulse and $t = 2T$, $\tau = \tau_{\pi}/2$ for the $\pi/2$ pulse. Consequently, we obtain:
\begin{subequations}
\label{eq:app:U-terms-2-pulses-mse}
\begin{gather}
\left\langle 2 U_{gg}^{(k) \ast} U_{ge}^{(k)} \right\rangle =  
\left\langle \frac{\sin^3 2\lambda}{(1+r^2)^{3/2}} 
\left[1 - \frac{i r \tan\lambda}{\sqrt{1+r^2}} \right] e^{i\pi r}  \right\rangle \nonumber \\
\times e^{- i\phi_k} , \label{eq:app:U-terms-2-pulses-mse-a} \\
\left\langle |U_{gg}^{(k)}|^2 - |U_{eg}^{(k)}|^2 \right\rangle 
= \left\langle \left( \frac{1-r^2}{1+r^2} \sin^2 2\lambda - \cos^2 2\lambda \right) \right. \nonumber \\
\times \left. \left( \frac{1-r^2}{1+r^2} \sin^2 \lambda - \cos^2 \lambda \right) \right\rangle , 
\label{eq:app:U-terms-2-pulses-mse-b}
\end{gather}
\end{subequations}
where $\Omega_{\text{eff}} \tau_{\pi} = \pi$ and we used notation
\begin{equation}
r \equiv \frac{\delta'_{12}}{\Omega_{\text{eff}}} = - \frac{p K}{m \Omega_{\text{eff}}} , \quad
\lambda \equiv \frac{\pi}{4} \sqrt{1+r^2} .
\end{equation}
In deriving Eqs.~\eqref{eq:app:U-terms-2-pulses-mse}, we neglected all terms that include the factors 
shown in Eqs.~\eqref{eq:app:av-all}
since, as we showed above, these averages are integrals over very rapidly oscillating functions and therefore they are essentially zero.

It is easy to see that, for small $r$, the average in Eq.~\eqref{eq:app:U-terms-2-pulses-mse-b} scales as 
$\big\langle |U_{gg}^{(k)}|^2 - |U_{eg}^{(k)}|^2 \big\rangle \approx - ( 1 - \pi/4 ) \langle r^2 \rangle$,
where $\langle r^2 \rangle = (K \sigma_{\text{th}}/ m \Omega_{\text{eff}})^2$, and its contribution to $\braket{\Pi}$ [i.e., the second term in Eq.~\eqref{eq:app:Parity-EV-entangled-1}] scales as $(1 - \pi/4)^N \langle r^2 \rangle^N$. Since we are interested in the parameter regime, in which $\langle r^2 \rangle \ll 1$ and $N \gg 1$, the contribution of this term to $\braket{\Pi}$ can be safely neglected. Also, the imaginary part of the average in Eq.~\eqref{eq:app:U-terms-2-pulses-mse-a} is an integral over an odd function of $p$ and therefore it is equal to zero. Using these facts, we obtain:
\begin{equation}
\braket{\Pi} =  (1-\eta)^N \text{Re}\! \left[ 2 c_g^{\ast} c_e e^{-i \sum_{k=1}^N \phi_k } \right] ,
\label{eq:app:Parity-EV-entangled-mse-1}
\end{equation}
where 
\begin{equation}
\eta \equiv 1 - \left\langle \frac{\sin^3 2\lambda}{(1+r^2)^{3/2}} 
\left[\cos \pi r + \frac{r \tan\lambda \sin \pi r}{\sqrt{1+r^2}} \right]   \right\rangle .
\label{eq:app:eta-exact}
\end{equation}
In particular, for $c_g = c_e = 1/\sqrt{2}$, Eq.~\eqref{eq:app:Parity-EV-entangled-mse-1} yields
\begin{equation}
\braket{\Pi} = (1-\eta)^N \cos\left(  \sum_{k=1}^N \phi_k \right) .
\label{eq:app:Parity-EV-entangled-mse-2} 
\end{equation}

\section{Effect of corrections to the detuning}
\label{sec:app-corrections}

As described in Sec.~\ref{sec:Doppler-shift}, if laser frequency chirping is used to compensate the evolving Doppler shift due to the acceleration of the atom, the time-dependent Raman detuning for the $k$th atom is given by Eq.~\eqref{eq:delta12-at-resonance}, where $b$ is the frequency chirp rate and we assumed that the Raman frequency at $t = 0$ is at the resonance for $p_k = 0$, as given by Eq.~\eqref{eq:resonance-condition}. While in Sec.~\ref{sec:Doppler-shift} we neglected the terms $(b - K a_k) T$ and $2 (b - K a_k) T$ compared to $\sigma_{\mathrm{th}} K/m$, in this Appendix we will take these relatively small corrections to the detuning into account.
As usual, we assume that all atoms experience the same constant acceleration: $a_k = a\ \forall k$. Correspondingly, $\phi_k = \phi = (b - Ka) T^2\ \forall k$. We also define $b = (1 + \epsilon) K a$, so that $b - K a = \epsilon K a$ and $\phi = \epsilon K a T^2$.

Similarly to the derivation in Appendix~\ref{sec:app-derivation} above, the expectation value of the parity operator is given by Eq.~\eqref{eq:app:Parity-EV-entangled-1}, where the average is over the momentum distribution of the thermal state, given by Eq.~\eqref{eq:average-P}. Also, the evolution operator for the two-pulse sequence $\pi$--$\pi/2$ is given by Eq.~\eqref{eq:U-2-pulses-with-detunerr}, and the evolution operator for each of the pulses in Eq.~\eqref{eq:U-2-pulses-with-detunerr}, $U_{t}^{(k)}(\tau)$, is given by Eq.~\eqref{eq:U-general}, where $t = T$, $\tau = \tau_{\pi}$ for the $\pi$ pulse and $t = 2T$, $\tau = \tau_{\pi}/2$ for the $\pi/2$ pulse. However, now we take into account that the Raman detuning has a different value for each of the two pulses: $\delta'_1 \equiv \delta'_{12}(T) =  - p K/m + \epsilon K a T$ for the $\pi$ pulse and $\delta'_2 \equiv \delta'_{12}(2T) =  - p K/m + 2 \epsilon K a T$ for the $\pi/2$ pulse. Consequently, we obtain:
\begin{subequations}
\label{eq:app:U-terms-2-pulses-corr}
\begin{gather}
\left\langle 2 U_{gg}^{(k) \ast} U_{ge}^{(k)} \right\rangle =  
\left\langle S(r_1, r_2)\! \left[1 - \frac{i r_2 \tan\lambda_2}{\sqrt{1+r_2^2}} \right]\! e^{i\pi r_1}  \right\rangle  
e^{- i\phi} , \label{eq:app:U-terms-2-pulses-corr-a} \\
\left\langle |U_{gg}^{(k)}|^2 - |U_{eg}^{(k)}|^2 \right\rangle 
= \left\langle \left( \frac{1-r_1^2}{1+r_1^2} \sin^2 2\lambda_1 - \cos^2 2\lambda_1 \right) \right. \nonumber \\
\times \left. \left( \frac{1-r_2^2}{1+r_2^2} \sin^2 \lambda_2 - \cos^2 \lambda_2 \right) \right\rangle , 
\label{eq:app:U-terms-2-pulses-corr-b}
\end{gather}
\end{subequations}
where 
\begin{equation}
S(r_1, r_2) \equiv \frac{\sin^2 2\lambda_1}{1+r_1^2} \frac{\sin 2\lambda_2}{\sqrt{1+r_2^2}} ,
\end{equation}
$\Omega_{\text{eff}} \tau_{\pi} = \pi$, and we used notation
\begin{equation}
r_n \equiv \frac{\delta'_n}{\Omega_{\text{eff}}} = - \frac{p K}{m \Omega_{\text{eff}}} + \frac{n \epsilon K a T}{\Omega_{\text{eff}}}, \quad
\lambda_n \equiv \frac{\pi}{4} \sqrt{1+r_n^2}, 
\end{equation}
for $n = 1,2$.

Once again, it is easy to see that, for small $r_n$, the average in Eq.~\eqref{eq:app:U-terms-2-pulses-corr-b} scales as 
$\big\langle |U_{gg}^{(k)}|^2 - |U_{eg}^{(k)}|^2 \big\rangle \approx - ( 1 - \pi/4 ) \langle r_2^2 \rangle$,
where $\langle r_2^2 \rangle = (K \sigma_{\text{th}}/ m \Omega_{\text{eff}})^2 + (2 \epsilon K a T/ \Omega_{\text{eff}})^2$, and its contribution to $\braket{\Pi}$ [i.e., the second term in Eq.~\eqref{eq:app:Parity-EV-entangled-1}] scales as $(1 - \pi/4)^N \langle r_2^2 \rangle^{N}$. Since we are interested in the parameter regime, in which $\langle r_2^2 \rangle \ll 1$ and $N \gg 1$, the contribution of this term to $\braket{\Pi}$ can be safely neglected. 

The average in Eq.~\eqref{eq:app:U-terms-2-pulses-corr-a} can be rewritten as
\begin{equation}
\left\langle 2 U_{gg}^{(k) \ast} U_{ge}^{(k)} \right\rangle = (B_R + i B_I) e^{- i\phi} = |B| e^{i(\theta - \phi)} ,
\end{equation}
where
\begin{subequations}
\label{eq:A-corr}
\begin{align}
B_R & =  \left\langle S(r_1, r_2) \left[\cos \pi r_1 + \frac{r_2 \tan\lambda_2}{\sqrt{1+r_2^2}} \sin \pi r_1 \right] \right\rangle , \\
B_I & =  \left\langle S(r_1, r_2) \left[\sin \pi r_1 - \frac{r_2 \tan\lambda_2}{\sqrt{1+r_2^2}} \cos \pi r_1 \right] \right\rangle ,
\end{align}
\end{subequations}
$|B| = \sqrt{B_R^2 + B_I^2}$, and $\theta = \tan^{-1} (B_I / B_R)$. Consequently, with $c_g = c_e = 1/\sqrt{2}$, the parity expectation value is
\begin{equation}
\braket{\Pi} = (1-\eta_{\mathrm{tot}})^N \cos[  N (\phi-\theta) ] ,
\label{eq:app:Parity-EV-entangled-corr} 
\end{equation}
where $\eta_{\mathrm{tot}} = 1 - |B|$.

The values of $\eta_{\mathrm{tot}}$ and $\theta$ can be easily evaluated via numerical integration in Eqs.~\eqref{eq:A-corr}. However, it is instructive to obtain approximate expressions which are valid when the detuning errors are small. We expand the integrands in Eqs.~\eqref{eq:A-corr} in the powers of $r_1$ and $r_2$ up to the second order, to obtain:
\begin{align}
& \eta_{\mathrm{tot}} = \eta + \eta_{\epsilon} 
\approx \kappa \left(\frac{K \sigma_{\text{th}}}{m \Omega_{\text{eff}}} \right)^2 
+ \left(\frac{\epsilon K a T}{\Omega_{\text{eff}}} \right)^2 , \label{eq:eta-approx-corr} \\
& \theta \approx (\pi-2) \sqrt{\eta_{\epsilon}} \approx (\pi-2) \frac{\epsilon K a T}{\Omega_{\text{eff}}} . \label{eq:theta-approx-corr}
\end{align}
We have verified that these approximate expressions are in excellent agreement with results obtained via numerical integration in Eqs.~\eqref{eq:A-corr} for parameter ranges relevant for typical experimental conditions with $^{133}$Cs atoms.

The first term in Eq.~\eqref{eq:eta-approx-corr}, $\eta$, is the parameter that quantifies the error due to the initial momentum spread, given by Eq.~\eqref{eq:eta-general} and, approximately, by Eq.~\eqref{eq:eta-approx-1}. The second term in Eq.~\eqref{eq:eta-approx-corr}, $\eta_{\epsilon}$, is is the parameter that quantifies the error due to the time-dependent correction to the detuning. The ratio of these two error parameters is, approximately,
\begin{equation}
\frac{\eta_{\epsilon}}{\eta} \approx \frac{\epsilon^2}{\kappa} \left( \frac{m a T}{\sigma_{\text{th}}} \right)^2
= \epsilon^2 \frac{E_{\text{kin}}}{2 \kappa \braket{E_{\text{vib}}}} ,
\end{equation}
where $\braket{E_{\text{vib}}}$ is the average vibrational energy of the atom in the trap, given by Eq.~\eqref{eq:E-vib}, and $E_{\text{kin}} = 2 m a^2 T^2$ is the kinetic energy acquired by the atom from $t=0$ to $t=2T$ (i.e., during the interferometer operation) due to the action of the constant force $m a$. For a $^{133}$Cs atom initially trapped in a harmonic potential with frequency $\nu_{\mathrm{trap}} = 10$~kHz at temperature $0.1$~$\mu$K, which subsequently moves in the gravitational field with $a = 9.8$~m/s$^2$ and $T = 1$~ms, we find $\eta_{\epsilon}/\eta \approx 1.91 \epsilon^2$. This ratio can be made very small by choosing $\epsilon \ll 1$, which will result in $\eta_{\epsilon}$ being just a small correction to $\eta$. Note that the interferometer phase scales as $\phi \propto \epsilon T^2$, while $\eta_{\epsilon} \propto \epsilon^2 T^2$, so for any value of $T$ it is possible to choose a value of $\epsilon$ such that $\eta_{\epsilon}/\eta \ll 1$ without reducing $\phi$ too much.

The appearance of the phase $\theta$ in Eq.~\eqref{eq:app:Parity-EV-entangled-corr} does not affect the phase uncertainty but it results in a shift of the interference fringes, which needs to be taken into account for an accurate measurement of $\phi$. The ratio
\begin{equation}
\frac{\theta}{\phi} \approx \frac{\pi - 2}{\Omega_{\text{eff}} T}  
\end{equation}
is independent of $\epsilon$. Typically, $\theta \ll \phi$, for example, with $\Omega_{\text{eff}} \sim 2\pi \times 10^5$~Hz and $T \sim 10^{-3}$~s, we find $\theta/\phi \sim 10^{-3}$. This ratio can be further reduced by increasing $\Omega_{\text{eff}}$ (which is also useful for reducing $\eta$ and $\eta_{\epsilon}$) and/or $T$.

\bibliography{Entangled_AI_Noise}

\end{document}